\documentstyle[prc,aps,floats,epsfig,psfig]{revtex}
\def\gtorder{\mathrel{\raise.3ex\hbox{$>$}\mkern-14mu
             \lower0.6ex\hbox{$\sim$}}}
\begin{document}
%\preprint{\vbox{\hbox{CERN-TH/2002-085}
%         \hbox{IFIC/02-17}}
\title{If sterile neutrinos exist, how can one determine
the total solar neutrino fluxes?}
\author{John N. Bahcall$^1$\thanks{jnb@ias.edu},
M.~C.~Gonzalez-Garcia$^{2,3,4}$
\thanks{concepcion.gonzalez-garcia@cern.ch} and  C.~Pe\~na-Garay$^{3}$
\thanks{penya@ific.uv.es}}
\vskip 1cm
\address{
  $^1$ School of Natural Sciences, Institute for Advanced Study, Princeton,
  NJ 08540\\
  $^2$ Theory Division, CERN, CH-1211 Geneva 23, Switzerland\\
  $^3$Instituto de F\'{\i}sica Corpuscular,
  Universitat de  Val\`encia -- C.S.I.C.\\
  Edificio Institutos de Paterna, Apt 22085, 46071 Val\`encia, Spain\\
  $^4$ C.N. Yang Institute for Theoretical Physics\\
  State University of New York at Stony Brook\\
  Stony Brook,NY 11794-3840, USA\\}
\maketitle
\vskip 1cm
\begin{abstract}
The $^8$B solar neutrino flux inferred from a global analysis of
solar neutrino experiments is within $11$\% ($1\sigma$) of the
predicted standard solar model value if only active neutrinos
exist, but could be as large as $1.7$ times the standard
prediction if sterile neutrinos exist. We show that the total
$^8$B neutrino flux (active plus sterile neutrinos) can be
determined experimentally to about  $10$\% ($1\sigma$) by
combining charged current measurements made with the KamLAND
reactor experiment and with the SNO CC solar neutrino experiment,
provided the LMA neutrino oscillation solution is correct and the
simulated performance of KamLAND is valid. Including also SNO NC
data, the sterile component of the $^8$B neutrino flux can be
measured by this method to an accuracy of about $ 12$\%
($1\sigma$) of the standard solar model flux.  Combining
Super-Kamiokande and KamLAND measurements and assuming the
oscillations occur only among active neutrinos, the $^8$B neutrino
flux can be measured to $6$\% ($1\sigma$); the total flux can be
measured to an accuracy of about $9$\%. The total $^7$Be solar
neutrino flux can be determined to an accuracy of about $28$\%
($1\sigma$) by combining measurements made with the KamLAND, SNO,
and gallium neutrino experiments. One can determine the total
$^7$Be neutrino flux to a $1\sigma$ accuracy of about $11$\% or
better by comparing data from the KamLAND experiment and the
BOREXINO solar neutrino experiment provided both detectors work as
expected. The $pp$ neutrino flux can be determined to about $15$\%
using data from the gallium, KamLAND, BOREXINO, and SNO
experiments.
\end{abstract}
%\end{titlepage}

\section{Introduction}
\label{sec:introduction}

We describe in this paper analysis procedures that can answer two
of the most important questions of neutrino research. How can one
determine the total solar neutrino fluxes (${\rm ^8B, ^7Be}$, and
$pp$) for comparison with solar model predictions? How can one
determine the sterile contribution to the total solar neutrino
fluxes? Our answers allow for the possibility of an arbitrary
mixture in solar neutrino oscillations of active and sterile
neutrinos, but require the correctness of the LMA solution of the
solar neutrino problems and careful attention to all the sources
of error (theoretical as well as experimental) \footnote{This
paper was originally written and posted on the electronic archive
(hep-ph) before the announcements of the recent SNO
results~\cite{snonc,snodn} and the improved SAGE measurement of
the gallium rate~\cite{sage2002} and also before our paper was
submitted for publication. We have included in the analysis
reported in this version of the paper, which we are submitting for
publication, the recent SNO and SAGE measurements. The ideas with
respect to the $^7$Be and $^8$B neutrinos are unchanged and the
numerical results have not been affected significantly, but the
present version is more up-to-date with respect to the input data.
We have also added, inspired by the SAGE discussion, a detailed
analysis of what one can learn about $pp$ neutrinos before there
is a dedicated experiment to measure just the $pp$ neutrino
flux.}.

We focus first on determining total fluxes by comparing charged
current (CC) observables measured in different experiments; this method
yields results as independent as possible of uncertainties due to the
presence of sterile neutrinos. We then describe how similar techniques
can be applied to determine total solar neutrino fluxes using a CC
experiment plus a neutrino-electron scattering experiment [or a
neutral current (NC) measurement], which yields results that depend
more on the sterile neutrino mix but which can nevertheless be
relatively accurate.

The numerical values we estimate for the expected precision with
which different quantities can be measured rely upon simulations
of the performance of the relevant experiments. Therefore the
accuracies that we quote are illustrative; the actual accuracies
that are obtainable can only be determined once the experimental
uncertainties are known.

\subsection{Flavor changes occur}\label{subsec:flavorchanges}

Neutrinos change flavors as they travel to the Earth from the
center of the Sun.  This flavor change was seen directly by the
comparison of the Sudbury Neutrino Observatory (SNO)
measurement~\cite{sno} of the charged current reaction for $^8$B
solar neutrinos with the Super-Kamiokande
measurement~\cite{sksol00} of the neutrino-electron scattering
rate (charged plus neutral current).  Even more clearly, flavor
change has been demonstrated by comparing the neutral current
measurement by SNO with the SNO CC measurement~\cite{snonc}.  The
conclusion that flavor changes occur among solar neutrinos, if
based solely upon the comparison of the SNO and Super-Kamiokande
event rates, is valid statistically at about the $3.2\sigma$
confidence level~\cite{sno,lisi,giunti,fiorentini,bahcall01}.  The
neutral current measurement of SNO increases the significance
level for flavor changes among solar neutrinos to the $5.3\sigma$
confidence level.

The SNO and Super-Kamiokande results demonstrate simply that new
physics is required to resolve the long-standing solar neutrino
problem~\cite{teo}, i.e., to understand the origin of the
discrepancy between the predictions of the standard solar
model~\cite{BP01} and the observed solar neutrino event rates
\cite{sno,sksol00,chlorine,kamiokande,sage,gallex,gno}. If one
includes the results of the Chlorine \cite{chlorine}, Kamiokande
\cite{kamiokande}, SAGE \cite{sage}, GALLEX \cite{gallex}, and GNO \cite{gno}
experiments together with the  SNO (CC) and Super-Kamiokande results,
then the combined measurements require~\cite{bahcall01} new
physics at $4.0\sigma$ and, if the relative temperature scaling of
the $^7$Be and $^8$B neutrino production reactions is taken into
account, at $7.4\sigma$. Helioseismological measurements confirm
the predicted sound speeds of the Standard Solar Model to better
than $0.1$\% and show that stellar physics cannot account for the
discrepancies between standard predictions and the observed solar
neutrino rates~\cite{helioseismology}.

\subsection{Current knowledge of the $^8$B solar neutrino flux if
only active neutrinos exist}\label{subsec:b8currentknowledge}

The combination of the charged current (CC)
and the charged plus neutral-current measurement with
Super-Kamiokande has been used by several
groups~\cite{sno,lisi,giunti,fiorentini} to determine the flux of
active  $^8$B neutrinos independent of the solar model. These
model-independent determinations of the active flux  exploit the
similarity between the response functions in the SNO and
Super-Kamiokande detectors~\cite{sno,lisi,rosen,villante}. In
addition, if one includes all the experimental data in a global
oscillation solution in which the $^8$B flux is a free parameter,
one obtains a similar (but slightly smaller) allowed range for the
$^8$B neutrino flux~\cite{BGP1,KS,BGP2}.
 The measurement of the NC rate by SNO~\cite{snonc} provides an
independent determination of the active $^8$B neutrino flux.

All of the analyses yield the same result: if electron neutrinos
oscillate into only active neutrinos, then the total $^8$B
neutrino flux is in excellent agreement with the flux predicted by
the Standard Solar Model.

This close agreement of the active $^8$B neutrino flux with the
total flux predicted by the Standard Solar Model
 (SSM)~\cite{BP01,BGP2} is, if the flux of sterile neutrinos is
small, an important confirmation of the quantitative theory of
stellar evolution.
 We summarize below the current best-estimates
and the associated $1\sigma$ uncertainties for the active $^8$B
neutrino flux, $\phi_{\rm active} ({\rm ^8B})$.
\begin{itemize}
\item
Standard solar model (BP00) prediction~\cite{BP01}:
\footnote{The recently measured low-energy cross section factor
reported by Junghans {\it et al.}~\cite{junghans01} is currently being reinvestigated.}
\begin{equation}
 \phi_{\rm active}
({\rm ^8B}) = 5.05 \times 10^6 {\rm
cm^2s^{-1}}(1^{+0.20}_{-0.16}).
\label{eq:itemfb8ssm}
\end{equation}
\item
Comparison of SNO and Super-Kamiokande event rates (cf. esp. Ref.~
\cite{sno}):
\begin{equation}
\phi_{\rm active} ({\rm ^8B}) = 5.44\times 10^6 {\rm
cm^2s^{-1}}(1\pm 0.18).
\label{eq:itemb8snosk}
\end{equation}
\item SNO neutral current measurement (assuming undistorted spectrum)~\cite{snonc}.
\begin{equation}
\phi_{\rm active} ({\rm ^8B}) = 5.09\times 10^6 {\rm
cm^2s^{-1}}(1\pm 0.12).
\label{eq:itemb8snonc}
\end{equation}
\item
Global neutrino oscillation analysis~(see
Table~\ref{tab:fbranges} this paper and Ref.~\cite{BGP2}):
\begin{equation}
 \phi_{\rm
active} ({\rm ^8B}) = 5.40\times 10^6 {\rm
cm^2s^{-1}}(1\pm 0.075). \label{eq:itemfb8global}
\end{equation}
\end{itemize}

The agreement, summarized in
 Eqs.~(\ref{eq:itemfb8ssm})--(\ref{eq:itemfb8global}), between the
SSM calculated flux and the measured active flux suggests that the
sterile neutrino contribution to the $^8$B neutrino flux may be
small. In this paper, we ignore this tempting suggestion and
instead concentrate on developing methods to determine
experimentally the total $^8$B and $^7$Be neutrino fluxes emitted
by the Sun, independent of the active-sterile mixture~(for a
discussion of earlier investigations of sterile neutrinos see
Refs.~\cite{bilenky,conchaandyossi}).

If we want to understand the particle physics implications of
solar neutrino research, we must determine if sterile neutrinos
are present in the solar neutrino flux . Moreover, the
original--and still valid--goal of solar neutrino research
was~\cite{teo} to compare solar model predicted and experimentally
measured (total) solar neutrino fluxes.

\subsection{What can one do if sterile neutrinos exist?}
\label{subsec:statusifsteriles}

What is the situation if sterile neutrinos exist? The total flux
of $^8$B neutrinos could in this case be much larger than the
standard solar model prediction; a major fraction of the total
flux that reaches the Earth could arrive in a form that is not
detected in solar neutrino experiments. The existing data
disfavor (at $5.4\sigma$ CL) oscillation into purely sterile
neutrinos. Nevertheless, a large sterile component is allowed
~\cite{BGP2,barger2001} if oscillations occur into a combination
of active and sterile neutrino states (see
Refs.~\cite{activesterile,four} for a description of the formalism
adopted here). The flux of sterile neutrinos could in principle
be large enough to destroy the apparently excellent agreement
between the flux predicted by the SSM and the true flux of $^8$B
neutrinos [which is assumed to be pure active neutrinos in the
comparison shown above in
Eqs.~(\ref{eq:itemfb8ssm})-(\ref{eq:itemfb8global})].

A measurement of the total $^8$B solar neutrino flux, including
the sterile component (if any),  will provide information that is
important for astrophysics and  for particle physics. The
motivation for investigating sterile neutrinos is not dependent
upon the LSND~\cite{lsnd} results that might suggest the existence
of sterile neutrinos.
Of course, the LSND
results are not supported by other experimental results (cf.
Ref.~\cite{karmen}) nor by theoretical predictions that have been
precisely confirmed (like the helioseismological verifications of
the standard solar model), as is the case for the inference of
flavor change based upon the SNO-Super-Kamiokande comparison.

Fortunately, the KamLAND reactor neutrino
experiment~\cite{kamland}, when combined with the SNO measurement
of the CC flux, is capable of providing a precise determination of
the total, i.e. the active plus the sterile, $^8$B neutrino flux.
We assume throughout this paper the correctness of the currently
favored large mixing angle (LMA) solution to the solar neutrino
problem. If the LMA solution is not correct, then all global
analyses of the available solar and reactor data indicate that
either the mass difference, $\Delta m^2$, or the vacuum mixing
angle, $\theta$, will be too small to produce a measurable effect
in the KamLAND experiment (see e.g.
Refs.~\cite{lisi,giunti,fiorentini,BGP1,KS,kamland,goswami}). In
this case, KamLAND will be unable to provide the information
required to determine the total $^8$B neutrino flux.

Here is the basic physical idea of the method we propose for
measuring the total $^8$B solar neutrino flux. For the
KamLAND~\cite{kamland} experiment, one will know accurately the
flux of anti-neutrinos from the $17$ reactors that contribute
significantly to the measured anti-neutrino events. From
measurements of the total event rate and the energy spectrum
induced by the surviving $\bar{\nu_e}$, the KamLAND
experimentalists can determine with
precision~\cite{kamland,BMW-KAM,BS-KAM,MP-KAM} the anti-neutrino
propagation parameters, $\Delta \bar m^2$ and $\tan^2 \bar
\theta$. Both the KamLAND measurement and the CC SNO measurements
are disappearance experiments for neutrinos (or anti-neutrinos) of
similar energies. For the CC measurement made with SNO, one does
not know the total $^8$B neutrino flux created in the Sun. But,
assuming conservation of CPT, one can use the propagation
parameters $\Delta \bar{m^2}$ and $\tan^2 \bar{\theta}$ determined
by KamLAND and the measured (by SNO) CC rate to solve for the flux
that gives the observed result. Summarizing, for the KamLAND
experiment one knows the total flux but not the propagation
parameters, which are measured. For the SNO CC experiment, one
will know (from KamLAND) the propagation parameters and therefore
can measure the total flux\footnote{The method described here is
of course more general than the specific application to the
KamLAND and SNO experiments. In order to determine the total flux,
it is sufficient that one measure a set of observables that do not
depend upon the solar neutrino flux (in this paper, the measured
quantities in the KamLAND experiment) and a quantity that does
depend upon the solar neutrino flux (here, the CC rate in SNO).}.

\begin{figure}[!ht]
\centerline{\psfig{file=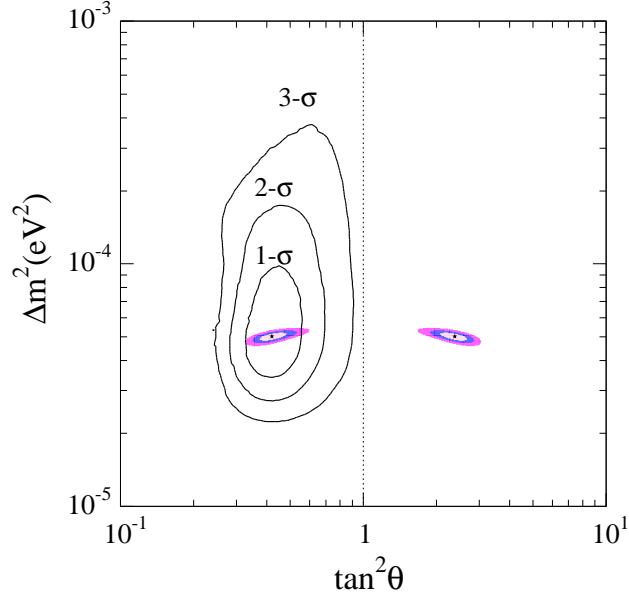,width=0.5\textwidth}}
\caption[]{Solar neutrino allowed region compared with simulated
KamLAND allowed region. The figure shows the currently allowed regions
of the solar neutrino oscillation parameters; the contours of equal CL
are labeled at $1\sigma$, $2\sigma$ and $3\sigma$.  This global
solution was obtained  assuming
pure active neutrino oscillations and using all the available solar and reactor data.
We include the recent SNO results~\cite{snonc,snodn}. The rates from the GALLEX/GNO~\cite{gallex,gno} and
SAGE~\cite{sage2002,sage} experiments have been averaged to provide a unique data
point ($72.4 \pm 4.7$ SNU). Some technical improvements
regarding neutrino cross sections and correlations of errors were
included in the analysis (see the Appendix). The two much smaller
allowed regions, placed symmetrically with respect to the line at
$\tan^2 \theta = 1$, represent the allowed regions, at
$1\sigma$,$2\sigma$, and $3\sigma$, that are obtained from a
simulation of what may be achievable with the KamLAND reactor
experiment. The best-fit point for the KamLAND simulation is assumed
to be the same as the best-fit point for the global solar neutrino
oscillation solution, namely, purely active neutrinos with: $\Delta
m^2 = 5.0\times 10^{-5} {\rm eV^2}$, $\tan^2 \theta = 0.42$.
\label{fig:solarvskamlandreal}}
\end{figure}

Figure~\ref{fig:solarvskamlandreal} shows the results of a refined
global solution for the solar neutrino oscillation parameters that
was made (see Sec.~\ref{sec:presentknowledge} and the
Appendix) using all the available solar and reactor data. The
figure displays the allowed solar neutrino oscillation contours at
$1\sigma$, $2\sigma$, and $3\sigma$.
 The results are obtained by
the procedures described most recently in Ref.~\cite{BGP2}, where
we have used in the present paper the analysis strategy (a) (of
Ref.~\cite{newbgp}) including the 1496 day Super-Kamiokande data
sample~\cite{smy} as well as the SNO CC, NC  and day-night observations
\cite{snonc,snodn}. We have also made some improvements (see
the Appendix) in the treatment of the uncertainties in the
neutrino cross sections and in the correlation of errors.

For comparison, we also show in Fig.~\ref{fig:solarvskamlandreal}
the small size of the expected allowed region for KamLAND if this
reactor anti-neutrino experiment observes a signal corresponding
to the current best-fit point of the solar neutrino analysis. In
calculating the KamLAND allowed region, we have made a
conservative estimate (see Sec.~\ref{subsec:precisionb8}),
following the principles discussed in Refs.~\cite{dgp,cc3}.
Figure~\ref{fig:solarvskamlandreal} shows clearly that KamLAND has
the potential for making a precise measurement of the solar
neutrino oscillation parameters, provided that the LMA is the
correct oscillation solution.

 The only  complication
involved in determining the total $^8$B flux from a comparison of
the KamLAND and SNO CC measurements results from the fact that for
the favored Large Mixing Angle (LMA) solar neutrino oscillation
solution matter effects in the Sun and the Earth can be
significant.  Matter effects are unimportant for the KamLAND
reactor experiment. The role of matter effects in solar neutrino
experiments depends somewhat upon the {\it a priori} unknown
active-sterile mixture, which introduces a calculable uncertainty
in the inferred total $^8$B neutrino flux.

In summary, the combination of the  SNO CC neutrino measurement
and the KamLAND anti-neutrino measurement will determine the total
 (active plus sterile) flux, $\phi_{\rm total} ({\rm ^8B})$, of
$^8$B solar neutrinos. By subtracting the previously determined
flux (see above), $\phi_{\rm active} ({\rm ^8B})$, from $\phi_{\rm
total} ({\rm ^8B})$, one can determine the flux of sterile solar
neutrinos.

Using similar reasoning, we shall also show that the combined
Super-Kamiokande and KamLAND measurements can be analyzed to yield
an accurate value for the total $^8$B flux, although in this case
the results are somewhat more sensitive to the active-sterile
admixture.

\subsection{$^7$Be solar neutrinos}
\label{subsec:be7solarneutrinos}

The flux of $^7$Be solar neutrinos, which in the SSM is
predicted\cite{BP01} to be $\phi({\rm ^7Be}_{\rm
active})~=~4.77\times 10^9 {\rm cm^2s^{-1}}(1\pm 0.10)$, can be
determined in a model independent way from a global analysis of
the solar neutrino data assuming only active neutrino
oscillations. For example, the latest analysis by Garzelli and
Giunti~\cite{giunti} yields $0.02\leq \phi({\rm ^7Be})/\phi({\rm
^7Be,SSM}) ~\leq~ 1.15$ at 99\% CL.

We shall also show in this paper that one can extract the value of
the $^7$Be neutrino flux from measurements of the gallium solar
neutrino experiments, GALLEX, SAGE, and GNO and the results of the
SNO and KamLAND measurements. The value of the $^7$Be flux that
will be derived in this way is relatively insensitive to the
assumed neutrino oscillation parameters, although it does depend
somewhat on the assumed contributions of the CNO, $pep$, and
$pp$ neutrino fluxes which we adopt from the standard solar
model. The constraint provided by the Chlorine experiment is not
very useful for determining the total $^7$Be neutrino flux.

One can obtain an independent measurement of the total $^7$Be
solar neutrino flux by comparing data from the KamLAND experiment
with data from the BOREXINO ~\cite{borexino} solar neutrino
experiment. If both the KamLAND and the BOREXINO detectors work as
expected, then this method will be more accurate than the methods
involving the gallium and chlorine radiochemical detectors.

\subsection{Appendix: just for aficionados}
\label{subsec:aficionados}

The determination of the total solar neutrino fluxes, and even
more so the determination of the sterile components of these
neutrino fluxes, requires precision in both the experimental
measurements and the theoretical calculations and analyses. We
present in the Appendix a refined discussion of the theoretical
errors, and their correlations, for the absorption cross for the
gallium and Chlorine solar neutrino experiments.

\subsection{Outline and suggested reading
strategy}\label{subsec:outline}

The outline of this paper is as follows. In
Sec.~\ref{sec:presentknowledge}, we describe the current
experimental knowledge of the $^8$B solar neutrino flux. Our
results are summarized in Table~\ref{tab:fbranges} both for the
special case of oscillations between purely active neutrinos and
for the general case of oscillations between electron neutrinos
and an active-sterile neutrino admixture. We limit our analysis to
the allowed LMA region of solar neutrino oscillations. We show in
Sec.~\ref{sec:determiningb8total} how one can use the CC
measurements with SNO and KamLAND to determine an accurate total
$^8$B solar neutrino flux including experimental and theoretical
uncertainties and the possibility of an appreciable active-sterile
admixture. We switch to the $^7$Be flux in Sec.~\ref{sec:be7} and
evaluate how well one can determine the total $^7$Be solar
neutrino flux by also using the results of the gallium (GALLEX,
SAGE, and GNO) solar neutrino experiments or the Chlorine
experiment. In Sec.~\ref{sec:klandes}, we investigate how well the
total $^8$B and $^7$Be neutrino fluxes can be determined using the
combined measurements of KamLAND and $\nu-e$ scattering observed
in the Super-Kamiokande (Sec.~\ref{subsec:kamlandsuperkb8}) and
BOREXINO (Sec.~\ref{subsec:kamlandborexino}) detectors. We show
that even in the presence of active-sterile admixtures the total
$^7$Be solar neutrino flux may be measured with relatively high
accuracy by comparing results from the KamLAND and the BOREXINO
experiments. We describe and analyze in
Sec.~\ref{sec:determiningpp} three strategies for determining the
total $pp$ solar neutrino flux in the absence of a dedicated
experiment that measures separately the $pp$ neutrinos. We
summarize and discuss our principal conclusions in
Sec.~\ref{sec:discuss}.

We urge the reader to turn first to Sec.~\ref{sec:discuss} and
read there the summary and discussion of our main results and
their implications. The rest of the paper can then be understood
more easily.

\section{Present knowledge of the $^8$B neutrino flux}
\label{sec:presentknowledge}

We generalize in this section the determination of the $^8$B
neutrino flux to the case in which $\nu_e$ oscillates into a
state that is a linear combination of active ($\nu_a$) and
sterile ($\nu_s$) neutrino states,

\begin{equation}
\nu_e \to \cos\eta \, \nu_x \,+\, \sin\eta \,\nu_s \, ,
\label{eq:linearcombination}
\end{equation}
where $\eta$ is the parameter that describes the active-sterile
admixture. This admixture arises naturally in the framework of
4-$\nu$ mixing~\cite{four}. The total $^8$B neutrino flux can be
written
\begin{equation}
\phi({\rm ^8B})_{\rm total} ~=~
\phi(\nu_e)~+~\phi(\nu_x)~+~\phi(\nu_{s}),
\end{equation}
where $\phi(\nu_{s})~=~\tan^2 \eta \times \phi(\nu_x)$. Clearly,
the larger the sterile component, the larger the value of
$\phi({\rm ^8B})_{\rm total}$ that is inferred from the
experimental data.

We have performed a  global analysis of the solar neutrino data
treating the total $^8$B  neutrino flux as a free parameter. The
details of the analysis procedure are the same as those used in
Ref.~\cite{BGP2} except where we explicitly state otherwise. We
concentrate here on the LMA region, $0.1<\tan^2\theta<10$,~
$10^{-5}<\Delta m^2/{\rm eV^2}<10^{-3}$.

 To take
account of the possibility of oscillations into sterile neutrinos,
we determine the allowed regions in the parameter space defined by
$\Delta m^2$, $\tan^2\theta$, and a third parameter, $\cos^2\eta$,
that is defined by Eq.~(\ref{eq:linearcombination}). It is
convenient to introduce the dimensionless parameter,
\begin{equation}
f_{\rm B}~=~\frac{\phi({\rm ^8B})_{\rm total}}{\phi({\rm ^8B})_{SSM}},
 \label{eq:defnfb}
\end{equation}
where $f_{\rm B}$ is the total $^8$B neutrino flux in units of the
predicted standard solar model flux.
\begin{table}[!b]
\centering \caption[]{The allowed range of the total $^8$B
neutrino flux.  The table presents the allowed range of $f_B$
[defined by Eq. (\ref{eq:defnfb})] that was found in a global
solution of all the currently available solar and reactor neutrino
data. The second column gives the allowed range of $f_B$ for an
arbitrary mixture of active and sterile neutrinos and the third
column gives the range for the case where only active neutrinos
are considered.  The results shown were obtained using
Eq.~(\ref{eq:fb8determination3}). \label{tab:fbranges}}
\begin{tabular}{ccc}
CL&$f_B$ (active + sterile)&$f_B$ (active)\\
\hline
$1\sigma$&0.99--1.25&0.99--1.15\\
$2\sigma$&0.92--1.47&0.92--1.22\\
$3\sigma$&0.84--1.67&0.84--1.29\\
\end{tabular}
\end{table}
The allowed range of $f_{\rm B}$ in the three dimensional space
of neutrino parameters, $\Delta m^2$, $\tan^2\theta$, and
$\cos^2\eta$, is determined by the equation
\begin{equation}
 \chi^2({\rm f_B}) ~\leq~ \chi^2_{\rm min} ~+~ \Delta \chi^2(1,{\rm
 CL}).
\label{eq:fb8determination3}
\end{equation}
Here $\Delta \chi^2(1,{\rm CL})$ is the change in $\chi^2$ that
corresponds to a specified confidence limit (CL) for one degree of
freedom. The computed values of $\chi^2$ are minimized for each
value of $f_{\rm B}$ with respect to $ \Delta m^2$,
$\tan^2\theta$, and $\cos^2\eta$.

Table~\ref{tab:fbranges} shows the currently allowed range for
$f_B$ for both the more general case where a mixture of active and
sterile neutrinos is assumed and for the more conventional case in
which only active neutrinos are considered. For purely active
neutrinos, the $1\sigma$ range is
\begin{equation}
f_{\rm B, \,active} ~=~1.07\pm 0.08~,
\label{eq:fbactive1sigma}
\end{equation}
and for an arbitrary mixture of active and sterile neutrinos, the
$1\sigma$ range is
\begin{equation}
f_{\rm B,\,active-sterile} ~=~1.07^{+0.18}_{-0.08}~.
\label{eq:fbactivesterile1sigma}
\end{equation}
The result shown earlier in
Eq.~(\ref{eq:itemfb8global}) for purely active neutrinos is taken
from Table~\ref{tab:fbranges}.

How does the possible existence of sterile neutrinos affect the
allowed range of $^8$B neutrino fluxes? We can calculate the
dependence of the allowed range of $f_B$ upon $\cos^2 \eta$ with
the aid of the inequality
\begin{equation}
\chi^2 (f_B,\cos^2\eta) ~\leq ~\chi^2_{\rm min} ~+~
\Delta\chi^2(2, {\rm CL}). \label{eq:fb8cos2eta}
\end{equation}
Here $\Delta \chi^2(2,{\rm CL})$ is the change in $\chi^2$ for a
specified CL  that corresponds to two degrees of freedom ;  the
computed values of $\chi^2$ are minimized at each point with
respect to $ \Delta m^2$, and $\tan^2\theta$.

\begin{figure}[!t]
\centerline{\psfig{file=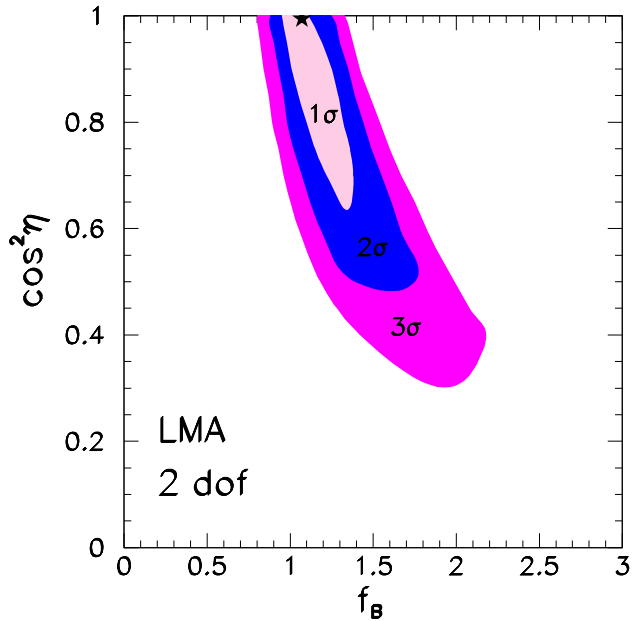,width=0.5\textwidth}}
\caption[]{The dependence of the inferred $^8$B flux on the
active--sterile admixture. The figure shows, as a function of the
active--sterile admixture, i.e.,  $\cos^2\eta$, the allowed range
of the $^8$B solar neutrino flux at $1\sigma$, $2\sigma$ and
$3\sigma$ CL. The star indicates the global best fit point for all
the currently available solar and reactor data; the star lies at
 $f_B = 1.07$ and $\eta = 0.0$ (purely active neutrinos).}
\label{fig:bosterile}
\end{figure}

Figure~\ref{fig:bosterile} shows the range of $f_{\rm B}$ as a
function of the active-sterile admixture, $\cos^2 \eta$, that is
obtained from Eq.~(\ref{eq:fb8cos2eta}) for the $1\sigma$,
$2\sigma$ and $3\sigma$ allowed regions. The allowed regions are
defined respect to the global minimum, which corresponds to purely
active oscillations with
 $\Delta m^2 = 5.0 \times 10^{-5}\, {\rm
eV^2}$, $\tan^2 \theta = 0.42$, and $f_{\rm B} = 1.07$.

Although pure sterile oscillations are forbidden at the 3~$\sigma$
CL (cf. Fig.~\ref{fig:bosterile}), a large sterile admixture in the
solar oscillations is still allowed. In fact, with the currently
available data, the largest allowed value at $3\sigma$ of the
sterile $^8$B neutrino flux corresponds to $f_{\rm B}= 2.2$ and
 $\cos^2 \eta = 0.3$ (for $2$ d.o.f.). For this extreme case, we
find
\begin{equation}
f_{\rm B, ~sterile~ max}~=~1.1,\label{eq:sterilemax}
\end{equation}

The quantity $f_{\rm B, ~sterile~ max}$ that appears in
Eq.~(\ref{eq:sterilemax}) is defined, analogous to $f_B$ in
Eq.~(\ref{eq:defnfb}), by the relation $f_{\rm B, ~sterile~ max} =
\phi({\rm ^8B})_{\rm sterile~ max}/\phi({\rm ^8B})_{\rm SSM}$. The
maximum value of $f_{\rm B, ~sterile~ max}$ is as large as the sum of
the active $^8$B neutrino fluxes ($f_x + f_e = 1.1 $, where $f_e =
0.348$, see Ref.~\cite{sno}) for this special case
\footnote{For 2 d.o.f, the maximum allowed value of $f_B$ is
 $2.2$ at $3\sigma$, but is $1.7$ for $1$ d.o.f, see
Table~\ref{tab:fbranges}. We have given in the Abstract the
maximum value for $1$ d.o.f.}.

What is the maximum allowed sterile contamination of the $^8$B
solar neutrino flux? Minimizing $\chi^2$ for the global solution
with respect to $\Delta m^2$, $\tan^2 \theta$, and $f_B$, we find
that the allowed range of $\cos^2 \eta$ satisfies
\begin{equation}
0.75 (0.40) \leq \cos^2 \eta \leq 1.0 \label{eq:etalimits}
\end{equation}
at $1\sigma({\rm or~} 3\sigma)$.

\section{How can we determine the total $^8$B neutrino flux using CC reactions?}
\label{sec:determiningb8total}

In this section, we will show how one can determine the
allowed range of the total $^8$B neutrino flux using the results
of the KamLAND reactor neutrino experiment and the SNO CC solar
neutrino experiment. We shall also estimate the accuracy with
which one can determine the total $^8$B flux.

Since we consider here only CC reactions that result from
disappearance experiments, the only difference between the role of
active neutrinos $\nu_{\mu}$ and $\nu_{\tau}$ and  sterile
neutrinos $\nu_{\rm sterile}$ arises from matter effects in the
Earth and in the Sun. Since sterile neutrinos do not interact with
matter, the effective potential for the $\nu_{\rm e}$--$\nu_{\rm
s}$ evolution in matter is $V_{\rm es}=V_{\rm e} - V_{\rm
s}=V_{\rm CC}+V_{\rm NC}$, since $V_s = 0$. The effective
potential $V_{es}$ is approximately half the potential for
$\nu_e$--$\nu_a$, $V_{\rm ea}=V_{\rm e}-V_{\rm a}=V_{\rm CC}$,
where $V_{\rm a}$ is the potential for the active neutrinos
$\nu_{\mu}$ and $\nu_{\tau}$. (The difference is exactly half for
a medium with equal number of neutrons, protons and electrons
because $V_{a}=V_{NC}= -G_F N_n/\sqrt{2}  \sim  -V_{CC}/2$ with
$V_{CC}=\sqrt{2} G_F N_e$.) We shall evaluate the expected
dependence of the inferred total $^8$B flux on the admixture of
sterile neutrinos [cf. Eq.~(\ref{eq:linearcombination})].

In Sec.~\ref{subsec:relationsfortotal} we present the formulae
that are used to determine the $^8$B flux with the aid of the
KamLAND and SNO CC experiments and in Sec.~\ref{subsec:admixture}
we illustrate the effect on the inferred $^8$B flux of the maximum
allowed (at $1\sigma$) sterile admixture. We estimate in
Sec.~\ref{subsec:precisionb8} the precision with which the $^8$B
flux can be determined including all the principal known sources
of uncertainties.

The reader who is interested in how well we can determine the
$^8$B flux,  but does not need to know the details of the
procedure, can get the main results by glancing at
Fig.~\ref{fig:bosterile} and Fig.~\ref{fig:fb8contours} and
Table~\ref{tab:fbo}.

% to here
In Sec.~\ref{subsec:kamlandsuperkb8}, we investigate how well the
total $^8$B flux can be determined using KamLAND in combination
with the $\nu-e$ scattering experiment, Super-Kamiokande.

\subsection{Relations that determine the total flux}
\label{subsec:relationsfortotal}

Suppose KamLAND observes a signal that  corresponds to LMA
$\overline\nu_e$ oscillations with parameters $(\Delta\bar
m^2,\tan\bar\theta^2)$. We assume the validity of the CPT theorem
so that constraints on anti-neutrino oscillation parameters
obtained from the KamLAND experiment apply to solar neutrino
experiments. We can then extract the $^8$B neutrino flux from the
following relation:
\begin{equation}
f_{\rm B}~=~ \frac{R^{\rm CC,exp}_{\rm SNO}} {R^{\rm SSM}_{\rm
SNO}}\times \frac{1} {\langle P_{ee} (\Delta
m^2,\tan^2\theta)\rangle_{\rm SNO}}, \label{fbodef}
\end{equation}
where
\begin{equation}
R^{\rm SSM}_{\rm SNO} = \int\! dE_\nu\, \phi^{\rm SSM} ({\rm ^8B},\,E_\nu) \sigma_e(E_\nu) =2.87
\, R^{\rm CC,exp}_{\rm SNO}
\label{eq:defnrssmsno}
\end{equation}
is the  CC rate for the SNO experiment\cite{sno} that is
predicted\cite{BP01,BGP2} by the standard solar model in the
absence of oscillations and ${\langle P_{ee} (\Delta
m^2,\tan^2\theta)\rangle_{\rm SNO}}$ is the average survival
probability for electron-flavor neutrinos created in the Sun.
Also, $E_\nu$ is the neutrino energy and $\sigma_{e}$ is the
weighted average $\nu_e$-d interaction cross-section, including
the experimental energy resolution function, ${\rm
Res} (T,\,T^\prime)$, where $T (T^\prime)$ is the measured (true)
recoil kinetic energy of the electron. Thus
\begin{equation}
\sigma_{e}(E_\nu)=\int_{T_{\rm min}}^{T_{\rm max}}\!dT
\int_0^{{T_{\rm max}}^\prime(E_\nu)}\!dT^\prime\,{\rm
Res}(T,\,T^\prime)\,
\frac{d\sigma_{e}(E_\nu,\,T^\prime)}{dT^\prime}\ .
\label{eq:sigmaccsno}
\end{equation}
The lower limit, $T_{\rm min}$, in the integral in
Eq.~(\ref{eq:sigmaccsno}) is taken here to be the threshold used by
the SNO Collaboration in Ref.~\cite{sno}
 ($T_{\rm min}=5$
MeV). The calculated value for
 the CC rate is not
sensitive to the assumed value of $T_{\rm max}$, as long as
$T_{\rm max} \geq 17$\,MeV.

The energy-averaged survival probability, $\langle P_{ee} (\Delta
m^2,\tan^2\theta)\rangle_{\rm SNO}$ for $\nu_e$  at SNO can be
computed using  the propagation parameters, $(\Delta \bar
m^2,\tan^2 \bar \theta)$ ,observed at KamLAND.  Thus
\begin{equation}
\langle P_{ee} (\Delta m^2,\tan^2\theta)\rangle_{\rm SNO}~=~
\frac{\int\! dE_\nu\, \phi^{\rm SSM}({\rm ^8B},\,E_\nu)
\sigma_e(E_\nu) P_{ee} (E_\nu,\Delta \bar m^2,\tan^2 \bar
\theta)} {R^{\rm SSM}_{\rm SNO}}. \label{eq:defnpee}
\end{equation}

\subsection{Illustrative dependence of total flux upon
active-sterile admixture} \label{subsec:admixture}

How much does a sterile neutrino admixture affect the inferred
total $^8$B neutrino flux? The dominant dependence on the sterile
admixture arises from matter effects within the Sun for larger
$\Delta m^2$ and within the Earth for smaller $\Delta m^2$ .

\begin{figure}[!t]
\centerline{\psfig{file=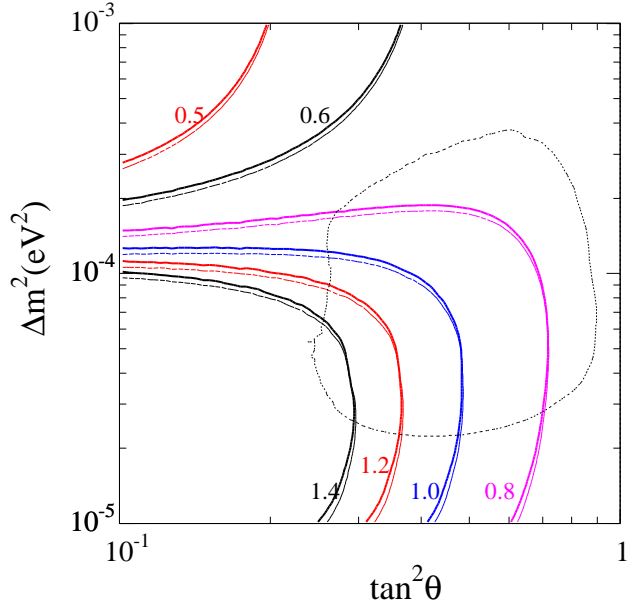,width=0.5\textwidth}}
\caption[]{Isocontours for the total $^8$B neutrino flux. The
figure compares isocontours for the $^8$B flux assuming purely
active neutrino oscillations (thicker lines) with the flux that
would be inferred for a $75$\% active-$25$\% sterile admixture
 (thinner lines). The results refer to a hypothetical comparison of
measurements from the KamLAND reactor experiment and the SNO CC
experiment.
We also show (dotted contour) the $3\sigma$ allowed region
obtained by a global fit to all of the allowed solar and reactor
data (cf. Fig.~\ref{fig:solarvskamlandreal}).}
\label{fig:fb8contours}
\end{figure}

 Figure~\ref{fig:fb8contours} shows the
isocontours of $f_{\rm B}$ in the LMA region for the pure active
case (thicker lines) and for the active-sterile case
 $\cos^2\eta=0.75$ (thinner lines). The isocontours are not very
different for the two cases. Moreover, the dependence on Earth
matter effects can be avoided experimentally by using only the
daytime CC measurement, once sufficient statistics are available.
The robustness of the inferred $^8$B solar neutrino flux can be
tested by comparing the $^8$B flux inferred using only daytime CC
measurements with the flux that is inferred when nighttime data
 (with corrections for Earth matter effects) are added to the
daytime data.

In Fig.~\ref{fig:fb8contours},we also show the $3\sigma$ LMA
contour (the dotted contour) obtained by the global analysis of
the solar neutrino data for the purely active case . Within the
$3\sigma$ LMA region, the maximum difference between the value of
$f_{\rm B}$ inferred allowing for possible sterile neutrinos and
the  value obtained assuming only active neutrino oscillations is
$+0.9$\% and $-3.5$\%. The dependence upon the active-sterile
mixture could become negligible if the correct $\Delta m^2$ lies
in the lower part of the LMA region and only daytime data is used
from the SNO CC measurements.

We conclude from Fig.~\ref{fig:fb8contours} that existence of
sterile neutrinos will not prevent an accurate measurement of the
total $^8$B neutrino flux.

\subsection{How accurately can the total $^8$B flux be determined?}
\label{subsec:precisionb8}

What is the overall precision expected in the determination of the
total $^8$B flux? From Eq.~(\ref{fbodef}) we can derive the
anticipated precision as
\begin{eqnarray}
\left(\frac{\sigma(f_{\rm B})} {f_{\rm B}}\right)^2&=&
\left(\frac{\sigma(R^{\rm CC,exp}_{\rm SNO})} {R^{\rm
CC,exp}_{\rm SNO}}\right)^2 + \left(\frac{\sigma(R^{\rm SSM}_{\rm
SNO})} {R^{\rm SSM}_{\rm SNO}}\right)^2 +
\left(\frac{\sigma(\langle P_{ee}\rangle_{\rm SNO,\,KamLAND})}
{\langle P_{ee}\rangle_{\rm SNO,\,KamLAND})}\right)^2\nonumber\\[+0.2cm]
&\equiv&{\delta(f_{\rm B})}^2_{\rm SNO,exp}+ {\delta(f_{\rm
B})}^2_{\rm SNO,Cross~Section}+{\delta(f_{\rm B})}^2_{\rm
SNO,\,KamLAND}, \label{eq:errorfbo}
\end{eqnarray}
where the $1\sigma$ errors are combined quadratically.

The current value of the first term in  Eq.~(\ref{eq:errorfbo}),
the $1\sigma$ uncertainty (statistical and systematic) of the
measured CC rate in SNO, is~\cite{snonc}
 ${\delta(f_{\rm B})}_{\rm
SNO,exp} = 6.15$\%. This uncertainty will undoubtedly decrease as
the results of analyzing more SNO CC data are reported.

The second term in Eq.~(\ref{eq:errorfbo}) represents the
uncertainty in the $\nu_e$-$^2$H absorption cross section. Much
progress has been made recently in evaluating this cross section,
see e.g. Refs.~\cite{deuteriumcs,deuteriumcsfieldtheory}, which
has led to an estimate $\sim 1$\% for the cross section
uncertainties other than radiative corrections. No definitive
calculation has yet been made of the radiative correction for the
CC reaction, but a reasonable estimate~\cite{deuteriumcs} is that
the cross sections given in Ref.~\cite{deuteriumcs} might be
increased by $2$\%. We adopt here a conservative uncertainty of
$\delta (f_{\rm B})_{\rm SNO,c.s.}= 2$\% ; the precise value chosen
for $\delta (f_{\rm B})_{\rm SNO,c.s.}$ is not very important at
this stage since other uncertainties are dominant. The $hep$
neutrino flux contribution to $\Delta f_B$ is negligible for our
purposes~\cite{sno,sksol00,BP01}

A detailed simulation is required to estimate $\sigma(\langle
P_{ee}\rangle_{\rm SNO,\,KamLAND})$, i.e., the uncertainty in the
average electron neutrino survival probability for  $^8$B solar
neutrinos observed in the SNO CC experiment as will be determined
by future KamLAND measurements. Here is how we estimate this
uncertainty. We generate the expected KamLAND signal for a fine
grid of points, $(\Delta \bar{m}^2,\, \tan^2\bar{\theta})$ that
spans the space of the allowed oscillation parameters determined
from solar neutrino experiments. For each grid point, we obtain
the allowed region by a $\chi^2$ analysis. We use statistical
errors corresponding to three years of data taking at KamLAND,
observing anti-neutrinos from reactors working at a constant
$78$\% of the maximal power. To be conservative, we also assumed a
neutrino energy threshold of $3.5$ MeV, in order to ensure that
the effects of natural radioactivity would be small. More details
on the KamLAND experiment can be found in Ref.~\cite{kamland};
details regarding the neutrino cross sections, statistical
procedures, and reactor fluxes used in the present paper are
described in Refs.~\cite{dgp,cc3}.

\begin{table}[!t]
\centering \caption[]{ Values of  $f_{\rm B}$ and associated
uncertainties obtainable from the SNO CC and KamLAND experiments.
The table presents the best-fit values for $f_{\rm B}$ (the total
$^8$B neutrino flux divided by the predicted standard solar model
$^8$B neutrino flux)  and associated uncertainties for a
representative set of possible oscillation parameters.  We have
used $\delta(f_{\rm B})_{\rm SNO,exp} = 6.15$\% and $\delta(f_{\rm
B})_{\rm SNO,CS}=2$\%. We consider active-sterile neutrino
admixtures permitted by the currently allowed global oscillation
solution [see Eq.~(\ref{eq:sterilemax})]. \label{tab:fbo}}
\begin{tabular}{ccccc}
\noalign{\smallskip}
 $\Delta\bar m^2$&$\tan^2\bar\theta$& $f_{\rm B}$ &
$\sigma(\langle P_{ee}\rangle_{\rm SNO,\,KamLAND})$&
${\rm Total}$ \\
&&&\%&\%\\
\noalign{\smallskip} \hline\noalign{\smallskip}
$5.0\times10^{-5}$ &$4.2\times10^{-1}$
& 1.09 &$^{+7.4}_{-6.9}$
& $^{+8.4}_{-8.1}$\\
\noalign{\medskip}
 $5.0\times10^{-5}$  &$5.01\times10^{-1}$ &
0.98 &$^{+7.2}_{-6.4}$
&$^{+8.3}_{-7.8}$\\
\noalign{\medskip}
 $5.0\times10^{-5}$  &$2.51\times10^{-1}$ &
1.51 &$^{+8.5}_{-8.5}$
&$^{+9.1}_{-8.9}$\\
\noalign{\medskip}
 $7.94\times10^{-5}$  &$4.2\times10^{-1}$ &
1.02 &$^{+6.6}_{-7.3}$
&$^{+9.0}_{-9.6}$\\
\noalign{\medskip}
 $7.94\times10^{-5}$  &$5.01\times10^{-1}$ &
0.94 &$^{+6.9}_{-9.2}$
&$^{+9.2}_{-11}$\\
\noalign{\medskip} $7.94\times10^{-5}$  &$2.51\times10^{-1}$ &
1.30 &$^{+7.1}_{-7.9}$
&$^{+9.4}_{-10}$\\
\noalign{\medskip}
 $3.16\times10^{-5}$  &$4.2\times10^{-1}$ &
1.01 &$^{+4.3}_{-5.3}$
&$^{+7.5}_{-8.2}$\\
\noalign{\medskip}
 $3.16\times10^{-5}$  &$5.01\times10^{-1}$ &
0.98 &$^{+6.1}_{-5.2}$
&$^{+8.7}_{-8.0}$\\
\noalign{\medskip}
 $3.16\times10^{-5}$  &$2.51\times10^{-1}$ &
1.57 &$^{+6.9}_{-8.1}$ &$^{+9.2}_{-10}$\\
\noalign{\smallskip}
\end{tabular}
\end{table}
In computing the inferred values of $f_{\rm B}$, we take account
of the fact that there could be a significant component of sterile
neutrinos in the incident $^8$B solar neutrino flux. We therefore
consider all active-sterile admixtures permitted by the global
oscillation solution shown in Fig.~\ref{fig:solarvskamlandreal}.
The numerical constraint on the currently allowed admixture is
given in Eq.~(\ref{eq:etalimits}).

 In principle,  for each simulated point there are two
allowed KamLAND  regions, one around $\Delta \bar{m}^2$ and
$\tan^2\bar{\theta}$ and another around $\Delta \bar{m}^2$ and
$\tan^2(\frac{\pi}{2}-\bar\theta)$. We discuss here only the range
of parameters within the first octant for the mixing angle since
global solar neutrino solutions
show~\cite{lisi,giunti,fiorentini,BGP1,KS,BGP2,goswami} that the
first octant is preferred. For each simulated KamLAND allowed
region centered on a specific $(\Delta \bar{m}^2,\,
\tan^2\bar{\theta})$, we compute $\sigma(\langle
P_{ee}\rangle_{\rm KamLAND})$. We repeated this procedure for a
grid of $81 \times 41$ points in the range
$0.1<\tan^2\bar{\theta}<10$, $10^{-5}<\Delta \bar{m}^2< 1 \times
10^{-3}$. The estimated uncertainty in $\langle P_{ee}\rangle_{\rm
SNO,\,KamLAND}$ varies with the grid point $(\Delta \bar{m}^2,\,
\tan^2\bar{\theta})$.

Table~\ref{tab:fbo} presents the best-fit value of $f_{\rm B}$,
the uncertainty in  $\sigma(\langle P_{ee}\rangle_{\rm
SNO,\,KamLAND})$, and the total expected uncertainty in inferring
$f_B$ from the combined SNO CC and the KamLAND neutrino reactor
measurements for a representative set of possible results for
$\Delta \bar{m}^2$ and $\tan^2\bar{\theta}$. In all the cases
shown in Table~\ref{tab:fbo}, we have used
  $\delta(f_{\rm B})_{\rm
SNO,exp} = 6.15$\% and $\delta(f{\rm B})_{\rm SNO,CS}=2$\% (see
previous discussion). The uncertainty contributed by the
possibility that sterile neutrinos exist, i.e., $\eta \not= 0$, is
rather modest; $\sigma(\langle P_{ee}\rangle_{\rm SNO,\,KamLAND})$
is typically reduced by $\sim 1$\%
 from the value shown in
Table~\ref{tab:fbo}.

\begin{figure}[!t]
\centerline{\psfig{file=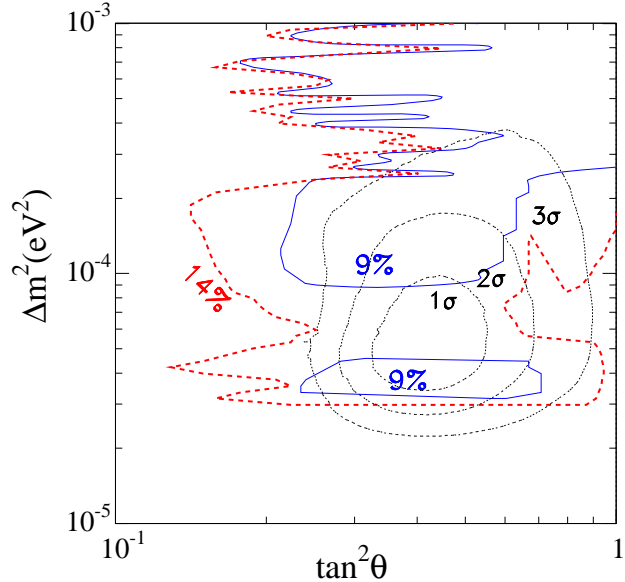,width=0.5\textwidth}}
\caption[]{Accuracy of determining the total $^8$B neutrino flux.
The figure displays  $1\sigma$ contours for the percentage
accuracy in determining the $^8$B flux that can be obtained from
the combined SNO CC and KamLAND data. The uncertainties were
calculated from Eq.~(\ref{eq:errorfbo}) and  the currently allowed
regions for the neutrino oscillation parameters were obtained by a
global fit to all of the allowed solar and reactor data (cf.
Fig.~\ref{fig:solarvskamlandreal}).} \label{fig:b8errors}
\end{figure}

 Figure~\ref{fig:b8errors} shows the $9$\% and the $14$\%
contours for the maximum percentage deviation from the best-fit
$f_{\rm B}$ value. To provide a context,  the figure also displays
the $1\sigma$, $2\sigma$ and $3\sigma$ allowed LMA regions
obtained by a global fit to the available solar and reactor data.
Within almost all of the current 1~$\sigma$ LMA allowed region,
the comparison of the KamLAND and the  SNO CC data will determine
the total $^8$B flux with an uncertainty that is less than
 $14$\%;
the uncertainty can be less than $10$\%  over a significant
fraction of the current $1\sigma$ allowed domain.  About $\sim$
 6\% of the current estimated uncertainty is due to the
experimental error in the SNO CC measurement, which hopefully will
be reduced as more CC data are accumulated.

For some purposes, it is convenient to know what is the average
expected accuracy in the determination of $f_{\rm B}$. We have
computed this average with the aid of a  Monte Carlo sampling of
the currently allowed solar neutrino oscillation parameters shown
in Fig.~\ref{fig:solarvskamlandreal}. The average positive and
negative uncertainties are approximately equal
 (cf. Table~\ref{tab:fbo}). We find an average $1\sigma$ uncertainty
of
\begin{equation}
\sigma(f_{\rm B}) ~=~9.6\%. \label{eq:averagefbsigma}
\end{equation}
A very significant component of $\sigma(f_{\rm B})$ comes from
$\sigma(\langle P_{ee}\rangle_{\rm SNO,\,KamLAND})$. We find,
averaged over the current $1\sigma$ solar neutrino oscillation
region,
\begin{equation}
\sigma(\langle P_{ee}\rangle_{\rm SNO,\,KamLAND}) ~=~7\%.
\label{eq:averagepeesigma}
\end{equation}

The sterile neutrino contribution to the $^8$B neutrino flux can
be determined by subtracting the active neutrino flux from the
total neutrino flux. Thus,
\begin{equation}
f_{\rm B,sterile}~=~ \frac{R^{\rm CC,exp}_{\rm SNO} -
R^{\rm NC,exp}_{\rm SNO}{\langle P_{ee} (\Delta
m^2,\tan^2\theta)\rangle_{\rm SNO}}}
{R^{\rm SSM}_{\rm
SNO} {\langle P_{ee} (\Delta
m^2,\tan^2\theta)\rangle_{\rm SNO}}}, \label{eq:fb8sterile}
\end{equation}
Using, as described above, the $1\sigma$ errors
for the total and the active fluxes of
$9.7$\% and $8$\% (assumed
for the SNO NC measurement), respectively, we estimate that the
sterile component of the $^8$B neutrino flux can be determined to
a precision of about $12.5$\%.

\section{How well can we determine the total $^7$B\lowercase{e} flux
using CC (radiochemical) experiments?}
\label{sec:be7}

We show in this section how the total $^7$Be solar neutrino flux
can be determined, with the judicial aid of other neutrino fluxes
predicted by the standard solar model~\cite{BP01}, by combining
the results of the GALLEX~\cite{gallex},GNO~\cite{gno}, and
SAGE~\cite{sage} gallium solar neutrino experiments with the
KamLAND and SNO CC measurements. We shall also explore the extent
to which the Chlorine
experiment~\cite{chlorine,chlorinehistorical} can provide
independent information about the $^7$Be solar neutrino flux.

We limit ourselves in this section to detectors that only observe
$\nu_e$ or $\bar \nu_e$, specifically, we consider here only the
radiochemical gallium and chlorine experiments and the reactor
anti-neutrino experiment, KamLAND. This limitation simplifies the
calculations with respect to the role of the sterile neutrinos.
However, the radiochemical experiments suffer from the
disadvantage of a lack of energy discrimination, which introduces
uncertainties involving the roles of the $pp$, $pep$, and CNO
neutrinos.

\begin{table}[!t]
\centering \caption[]{{\bf Gallium neutrino capture
rates and solar
neutrino fluxes.} The table  presents the predicted standard solar
model~\cite{BP01} neutrino fluxes and the calculated gallium
neutrino capture rates, with $1\sigma$ uncertainties from all
sources (combined quadratically). The neutrino fluxes are the same
as in the original BP00 model~\cite{BP01}.
The last column of the table presents the capture rate for gallium
predicted by the best fit LMA solution.
The total rates were calculated using the neutrino absorption
cross sections and their uncertainties that are given in
Ref.~\cite{GaCS}.
 \protect\label{tab:garates}}
\begin{tabular}{llcc}
\noalign{\bigskip} \hline \noalign{\smallskip}
Source&\multicolumn{1}{c}{Flux}&Ga (SSM)&Ga (LMA)\\
&\multicolumn{1}{c}{$\left(10^{10}\ {\rm cm^{-2}s^{-1}}\right)$}&(SNU)&(SNU)\\
\noalign{\smallskip} \hline \noalign{\smallskip}
$pp$&$5.95 \times 10^{0}~~\left(1.00^{+0.01}_{-0.01}\right)$&69.7&40.4\\
$pep$&$1.40 \times 10^{-2}\left(1.00^{+0.015}_{-0.015}\right)$&2.8&1.51\\
$hep$&$9.3 \times 10^{-7}$&0.1&0.023\\
${\rm ^7Be}$&$4.77 \times 10^{-1}\left(1.00^{+0.10}_{-0.10}\right)$&34.2&18.6\\
${\rm ^8B}$&$5.05 \times 10^{-4}\left(1.00^{+0.20}_{-0.16}\right)$&12.2&4.35\\
${\rm ^{13}N}$&$5.48 \times
10^{-2}\left(1.00^{+0.21}_{-0.17}\right)$&3.4&1.79\\
${\rm ^{15}O}$&$4.80 \times
10^{-2}\left(1.00^{+0.25}_{-0.19}\right)$&5.5&2.83\\
${\rm ^{17}F}$&$5.63 \times
10^{-4}\left(1.00^{+0.25}_{-0.25}\right)$&0.1&0.03\\
\noalign{\medskip}
&&\hrulefill&\hrulefill\\
Total&&$128^{+9}_{-7}$&$69.6$\\
\noalign{\smallskip}
\end{tabular}
\end{table}

We begin by describing in Sec.~\ref{subsec:be7procedure} the
general procedure for determining the $^7$Be neutrino flux.  We
then evaluate in Sec.~\ref{subsec:uncertaintiesbe7} the principal
sources of error, taking account of experimental and theoretical
uncertainties as well as the possibility of an appreciable sterile
neutrino component in the incident solar neutrino flux. We present
in Sec.~\ref{subsec:accuracybe7} the numerical results for the
uncertainties due to different factors and evaluate the overall
accuracy with which the total $^7$Be flux can be determined.

Using data from either the gallium or the Chlorine~\cite{chlorine}
experiments, the same procedure can be applied for inferring the
$^7$Be neutrino flux. For simplicity, we describe the procedure in
Sec.~\ref{subsec:be7procedure}--Sec.~\ref{subsec:accuracybe7} with
reference to the more promising case provided by the gallium
experiments. In Sec.~\ref{subsec:chlorine}, we investigate how
accurately one can determine the $^7$Be flux using data from the
Chlorine experiment instead of the gallium experiment.

In Sec.~\ref{subsec:kamlandborexino}, we use the techniques
developed in this section to explore the accuracy with which
KamLAND and BOREXINO can determine $f_{\rm Be}$.

\subsection{Procedure for determining the total $^7$Be solar
neutrino flux} \label{subsec:be7procedure}.

 Table~\ref{tab:garates} shows the neutrino fluxes and the
event rates in the gallium solar neutrino experiments that are
predicted by the standard solar model~\cite{BP01,BGP2} . The table
also shows the event rate predicted by the best-fit LMA solution.
From Table~\ref{tab:garates} it is clear that one must make a
strong assumption about the $pp$ neutrino flux in order to
determine the $^7$Be flux. One must also make assumptions
regarding the best-value and the uncertainties in the CNO and
$pep$ fluxes. This situation is different than the purely
empirical procedure described in Sec.~\ref{sec:determiningb8total}
for determining the $^8$B neutrino flux; the $^8$B solar neutrino
flux can be determined independent of all considerations regarding
the standard solar model.

We assume throughout this section the correctness of the
calculated standard solar model~\cite{BP01} values for the
neutrino fluxes and their uncertainties, except for the $^7$Be and
$^8$B fluxes which we want to determine from solar neutrino
experiments\footnote{The ultimate astronomical goal of solar
neutrino experiments is to determine all of the solar neutrino
fluxes directly from experiment, but there are too few
experimental constraints to make this possible at the present
time.}.

Fortunately, as we shall see, if we assume the SSM predictions and
their uncertainties for all but the $^7$Be and $^8$B fluxes, then
we can infer an interesting range for the total solar $^7$Be
neutrino flux if we use data from the gallium experiments. The
situation is less promising if we use data from the Chlorine
experiment rather than the gallium experiments.

Suppose that KamLAND observes a signal that corresponds to
$\overline\nu_e$  oscillations with parameters $(\Delta\bar
m^2,\tan\bar\theta^2)_{\rm KamLAND}$, then the  expected event
rate in the gallium experiments is a sum of the contributions from
the different neutrino fluxes, namely,
\begin{eqnarray}
R_{\rm Ga}&=& f_{\rm B}\, R^{\rm ^8B,SSM}_{\rm Ga} \langle
P_{ee}(\Delta \bar m^2,\tan^2\bar\theta)_{\rm KamLAND}\rangle^{\rm
^8B}_{\rm Ga} \,+\,f_{\rm Be} \, R^{\rm ^7Be,SSM}_{\rm Ga} \langle
P_{ee}(\Delta \bar m^2,\tan^2\bar\theta)_{\rm KamLAND}\rangle^{\rm
^7Be}_{\rm Ga}
\nonumber\\[+0.2cm]
&&+\sum_i f_i R^{\phi_i,SSM}_{\rm Ga} \langle P_{ee} (\Delta\bar
m^2,\tan^2\bar\theta)_{\rm KamLAND} \rangle^{\phi_i}_{\rm Ga} .
\label{eq:rga}
\end{eqnarray}

\begin{figure}[!t]
\centerline{\psfig{file=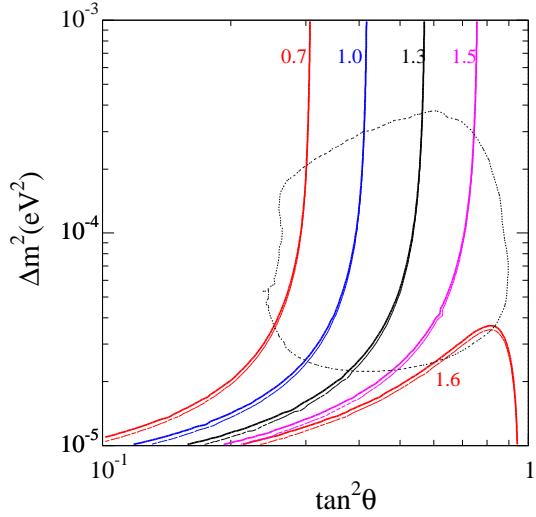,width=3in}}
\caption[]{Isocontours for the total $^7$Be flux. The figure
compares isocontours for the total $^7$Be flux assuming
oscillations between $\nu_e$ and purely active neutrinos (thicker
lines) or  oscillations between $\nu_e$ and a  75\% active-25\%
sterile admixture (thinner lines).  The results were obtained by
solving Eq.~(\ref{eq:fbe}) as described in the text. The
differences between the pure active contours and the $75$\%-$25$\%
admixtures are less than $2$\% within the currently allowed
$3\sigma$ solution space, which is shown by the dotted line in the
figure (cf. Fig.~\ref{fig:solarvskamlandreal}). The dotted curve
represents the $3\sigma$ allowed regions from the analysis of the
solar data.} \label{fig:be7contours}
\end{figure}

In the last term in Eq.~(\ref{eq:rga}), we include the
contributions from $hep$, $pep$, CNO and $pp$ neutrinos. By
analogy with Eq.~(\ref{eq:defnfb}), we have defined the factors
$f_i$ as the ratios between the ``true'' solar neutrino fluxes and
the fluxes predicted by the standard solar model.

We can solve Eq.~(\ref{eq:rga}) for the $^7$Be solar neutrino flux
as follows. We substitute into Eq.~(\ref{eq:rga}) the value of
$f_{\rm B}$ determined, independent of the solar model, from the
KamLAND and SNO CC measurements, as discussed in the previous
section (Sec.~\ref{sec:determiningb8total}). We also assume as
a first approximation that all the solar neutrino fluxes but the
$^8$B and $^7$Be fluxes are equal to the values predicted by the
SSM; we investigate later the accuracy of this approximation. With
these assumptions, we can then solve for $f_{\rm Be}$ by equating
 $R_{\rm Ga} = R^{\rm exp}_{\rm Ga}=72.4\pm 4.7$ SNU.
Thus\hfill\break
\vbox{
\begin{eqnarray}
f_{\rm Be}=\frac{1}{R^{\rm ^7Be,SSM}_{\rm Ga} \langle P_{ee}
(\Delta\bar m^2,\tan^2\bar\theta)_{\rm KamLAND} \rangle^{\rm
^7Be}_{\rm Ga}} &\Big[&R^{\rm exp}_{\rm Ga} - \sum_i R^{\rm
\phi_i,SSM}_{\rm Ga}
 \langle P_{ee} (\Delta\bar m^2,\tan^2\bar\theta)_{\rm KamLAND}
\rangle^{\phi_i}_{\rm Ga} \nonumber \\ [+0.2cm]  & & -R^{\rm
^8B,SSM}_{\rm Ga} \frac{R^{\rm CC,exp}_{\rm SNO}} {R^{\rm
CC,SSM}_{\rm SNO}}\frac{ \langle P_{ee} (\Delta\bar
m^2,\tan^2\bar\theta)_{\rm KamLAND}\rangle^{\rm ^8B}_{\rm Ga}}
{\langle P_{ee} (\Delta\bar m^2,\tan^2\bar\theta) _{\rm
KamLAND}\rangle_{\rm SNO}} \Big] \label{eq:fbe} .
\end{eqnarray}
}

Figure~\ref{fig:be7contours} shows that for the $^7$Be neutrino flux
the uncertainty due to the potential effect of sterile neutrinos is
small. The figure shows the isocontours of $f_{\rm Be}$ in the LMA
region for the case of purely active case (thicker lines) and for the
mixture of active and sterile neutrinos with $\cos^2\eta=0.75$
(thinner lines). Within the $1\sigma$ ($3\sigma$) parameter region,
the difference in the value of $f_{\rm B}$ between the two oscillation
scenarios is less than $1$\% ($2$\%).

\subsection{Principal sources of uncertainty in determining the $^7$Be total flux }
\label{subsec:uncertaintiesbe7}

The uncertainty in the inferred total $^7$Be solar neutrino flux
can be estimated from Eq.~(\ref{eq:fbe}). Including just the
largest contributions, we can write the fractional uncertainty in
the total $^7$Be neutrino flux as
\begin{eqnarray}
\left(\frac{\sigma(f_{\rm Be})} {f_{\rm Be}}\right)^2 &=&
{\delta(f_{\rm Be})}^2_{\rm Ga,exp} ~+~ {\delta(f_{\rm
Be})}^2_{\rm\,Ga,cross section}~+~ {\delta(f_{\rm Be})}^2_{\rm
KamLAND}~+~ {\delta(f_{\rm Be})}^2_{\rm CNO} +
 \nonumber \\
& &  ~+~ {\rm smaller~terms} \label{eq:errorsfbe}
\end{eqnarray}
where ${\delta(f_{\rm Be})}_{\rm Ga,exp}$ is the uncertainty from
the experimental error of the gallium rate, ${\delta(f_{\rm
Be})}_{\rm Ga, cross~section}$ is the theoretical uncertainty from
the calculated  $\nu_e$-Ga absorption cross sections \cite{GaCS},
${\delta(f_{\rm Be})}^2_{\rm KamLAND}$ is the uncertainty arising
from the allowed range of neutrino parameters determined by
KamLAND and SNO [which affects all of the averaged survival
probabilities in Eq.~(\ref{eq:fbe})], and  ${\delta(f_{\rm
Be})}^2_{\rm CNO}$ is the uncertainty due to the quoted errors in
the standard solar model calculation of the CNO fluxes (see
Table~\ref{tab:garates} and Ref.~\cite{BP01}). We have omitted
from Eq.~(\ref{eq:errorsfbe}) a number of sources of error that
contribute only relatively small uncertainties. The omitted
sources of error [and their range of contributed uncertainties for
a representative set of oscillation parameters] include the
experimental error in the SNO CC measurements [($1.2 \pm 0.6$)\%],
the uncertainty in the theoretical absorption cross section for
the SNO CC experiment [($0.7 \pm 0.2$)\%], and the uncertainty in
standard solar model calculation of the $pp$ neutrino flux
[($0.4 \pm 0.1$)\%].

We now discuss how we estimate the uncertainties in $f_{\rm Be}$.
Because they require special treatment, we first discuss the
uncertainty arising from the theoretical neutrino capture cross
sections for gallium and then discuss the uncertainty resulting
from the range of allowed neutrino parameters determined by
KamLAND\footnote{The SNO CC results are used to select the first
quadrant for $\theta$ (cf. Fig.~\ref{fig:solarvskamlandreal}),
but for brevity we refer to the range of neutrino parameters
determined by KamLAND.}.

To evaluate the uncertainties arising from the gallium absorption
cross sections for neutrino sources with continuous energy
distributions, we use Tables 2--4 of Ref.~\cite{GaCS}; these
tables give the best-estimate and the $\pm 3\sigma$ limits for the
theoretical cross sections. For the neutrino lines from $^7$Be and
$pep$, we have checked that the shape of the line~\cite{be7shape}
does not affect significantly the error estimate. Therefore, we
use for the $^7$Be and $pep$ lines the error estimates given in
Eq.~(41) and Eq.~(42) of Ref.~\cite{GaCS}. We have adopted the
conservative procedure described in Sec.~XII.A.4 of
Ref.~\cite{GaCS}, in which the uncertainties in all of the low
energy ($<2$ MeV) cross sections are fully correlated, while the
uncertainties for the ($^8$B) neutrinos above $2$ MeV are treated
separately. All of the cross sections for low energy neutrinos
move up or down together, reflecting the fact that the dominant
uncertainties for low energy neutrinos are common to all sources.
For higher-energy neutrinos, a number of excited states dominate
the calculated absorption cross section (For a more explicit
description of how the cross section errors are treated, see the
Appendix.).

To calculate the uncertainty associated with the range of allowed
neutrino parameters determined by KamLAND,  we first solve
Eq.~(\ref{eq:fbe}) with the appropriate average survival
probabilities computed for each neutrino oscillation point
 (${\Delta \bar m^2,\tan^2\bar\theta}$) in the allowed region (cf.
Fig.~\ref{fig:solarvskamlandreal}).  We consider active-sterile
neutrino admixtures permitted by the currently allowed global
oscillation solution [see Eq.~(\ref{eq:etalimits})]. The solution
of  Eq.~(\ref{eq:fbe}) determines $f_{\rm Be}({\rm best~fit})$ for
that particular point in oscillation parameter space. Then we
construct a $1\sigma$ allowed region, a set of points $({\Delta
m^2,\tan^2\theta})$, around the chosen point using the simulated
characteristics of the KamLAND experiment (see
Sec.~\ref{subsec:precisionb8} and Refs.~\cite{kamland,dgp,cc3}).
We define $\delta(f_{\rm Be})_{\rm KamLAND}$ for the chosen
(${\Delta \bar m^2,\tan^2\bar\theta}$) to be the maximum (or
minimum) value of $\left( f_{\rm Be}({\Delta
 m^2,\tan^2\theta}) - f_{\rm Be}({\rm best~fit})\right)/
f_{\rm Be}({\rm best~fit})$. In
 practice, the inclusion of sterile neutrinos only slightly
 affects the computed range of $\delta(f_{\rm Be})_{\rm
KamLAND}$.

For all other quantities, we estimate the associated uncertainty
at a particular point (${\Delta \bar m^2,\tan^2\bar\theta}$) in
the following way. Given a source of uncertainty $i$ (for example,
the measured capture rate for the gallium experiments) with
1$\sigma$ error $\sigma_i$, we obtain $\delta(f_{\rm Be})_{i}$
from the relation
\begin{equation}
\delta(f_{\rm Be})_{i} =\frac{f_{\rm Be}(i\pm\sigma_i)- f_{\rm
Be}(i)}{f_{\rm Be}(i)} . \label{eq:deltafbe}
\end{equation}
Here we denote by  $f_{\rm Be}(i)$ the value of $f_{\rm Be}$
obtained from Eq.~(\ref{eq:fbe}) when all the parameters are
assigned their best-estimate values. The quantity $f_{\rm
Be}(i\pm\sigma_i)$ is calculated from Eq.~(\ref{eq:fbe}) using the
best-estimate values of all variables except $i$; the variable $i$
is set equal to its best-fit value $\pm$ the corresponding
1$\sigma$ uncertainty. In calculating ${\delta(f_{\rm Be})}^2_{\rm
CNO}$, we shift the three CNO neutrino fluxes by $\pm 1\sigma$
simultaneously and in the same direction, reflecting the
correlation between the CNO fluxes in the standard solar model.

The uncertainty, $\delta(f_{\rm Be})_{i}$, that is calculated from
Eq.~(\ref{eq:deltafbe}) will in general depend upon the assumed
value for (${\Delta \bar m^2,\tan^2\bar\theta}$) within the
KamLAND allowed region.  This dependence persists even if
$\sigma_i$ is independent of (${\Delta \bar
m^2,\tan^2\bar\theta}$) (which is true, e.g., for the measured
gallium capture rate). In fact, the positive and negative values
for $\delta(f_{\rm Be})_{i}$ will generally not be equal.

\begin{table}[!ht]
\caption{Best-fit values of $f_{\rm Be}$ and associated
uncertainties (gallium based simulation). The table shows the
best-fit values and associated uncertainties that are obtained by
solving Eq.~(\ref{eq:fbe}) for a representative set of oscillation
parameters within the expected KamLAND and SNO allowed region (cf.
Fig.~\ref{fig:solarvskamlandreal}). \label{tab:fbe}}
\begin{tabular}{cccccccc}
& & & \multicolumn{5}{c} {Uncertainties $\delta(f_{\rm Be})_i$
(\%)}\\
\noalign{\smallskip}\cline{4-8}\noalign{\smallskip}$\Delta\bar
m^2$&$\tan^2\bar\theta$& $f_{\rm Be}$ &  ${\rm Ga, exp}$ & ${\rm
Ga, c.s.}$ &${\rm KamLAND}$ &  ${\rm CNO (flux)}$ &
${\rm Total}$ \\
\noalign{\smallskip}\hline\noalign{\smallskip} $5.0\times10^{-5}$
&$4.2\times10^{-1}$
& 1.15 &22
&$^{+~9}_{-15}$&$^{+~9}_{-10}$ &4.5&$^{+25}_{-28}$\\
\noalign{\medskip} $5.0\times10^{-5}$  &$5.01\times10^{-1}$ &
1.31 &20&
$^{+~9}_{-13}$&$^{+7}_{-8}$ &4.0&$^{+22}_{-25}$\\
\noalign{\medskip} $5.0\times10^{-5}$  &$2.51\times10^{-1}$ &
0.63 &24&
$^{+15}_{-22}$&$^{+17}_{-17}$ &8.1&$^{+42}_{-43}$\\
\noalign{\medskip} $7.94\times10^{-5}$  &$4.2\times10^{-1}$ &
1.09 &23&
$^{+10}_{-15}$&$^{+14}_{-13}$ &4.8&$^{+28}_{-31}$\\
\noalign{\medskip} $7.94\times10^{-5}$  &$5.01\times10^{-1}$ &
1.25 &20&
$^{+~9}_{-14}$&$^{+13}_{-11}$ &4.2&$^{+26}_{-27}$\\
\noalign{\medskip} $7.94\times10^{-5}$  &$2.51\times10^{-1}$ &
0.58 &36&
$^{+16}_{-24}$&$^{+23}_{-22}$ &9.1&$^{+47}_{-50}$\\
\noalign{\medskip} $3.16\times10^{-5}$  &$4.2\times10^{-1}$ &
1.26 &21&
$^{+~9}_{-14}$&$^{+8}_{-7}$ &4.0&$^{+25}_{-26}$\\
\noalign{\medskip} $3.16\times10^{-5}$  &$5.01\times10^{-1}$ &
1.40 &19&
$^{+~8}_{-13}$&$^{+6}_{-6}$ &3.6&$^{+22}_{-24}$\\
\noalign{\medskip} $3.16\times10^{-5}$  &$2.51\times10^{-1}$ &
0.73 &31& $^{+13}_{-16}$&$^{+21}_{-15}$ &6.6&$^{+41}_{-41}$
\end{tabular}
\end{table}

\subsection{The accuracy with which the total $^7$Be flux can be
determined} \label{subsec:accuracybe7}

Table~\ref{tab:fbe} presents the calculated uncertainties and the
best-fit values of $f_{\rm Be}$ for a representative set of
possible neutrino oscillation parameters, ($\Delta \bar{m}^2,
\tan^2\bar{\theta}$), that may be obtained from the KamLAND
measurements . The largest uncertainty, $\sim 22$\%, is due to the
experimental error on the measured gallium
rate~\cite{sage,gallex,gno}. The two next largest uncertainties,
both $\sim 12$\%, arise from the theoretical calculation of the
gallium absorption cross sections and the simulated errors in the
KamLAND measurements.  The rather large uncertainty due to the
gallium cross sections requires explanation since the
uncertainties on the individual cross sections are much
smaller~\cite{GaCS} [e.g., $2.3$\% for $pp$ neutrinos and $5$\%
 (average) for $^7$Be neutrinos]. The amplification in the error
due to the cross sections arises because all of the low energy
cross section errors are fully correlated. The gallium cross
section errors are added linearly in calculating $\delta f_{\rm
Be}$ [cf. Eq.~(\ref{eq:fbe})] rather than being combined
quadratically.

We have also computed a representative (average) error in the
determination of the total $^7$Be flux by a Monte Carlo sampling
of the allowed KamLAND oscillation region shown in
Fig.~\ref{fig:solarvskamlandreal}. We find
\begin{equation}
f_{\rm Be} ~=~f_{\rm Be,best~fit}\left[1.00 ^{+0.27}_{-0.29}
\right]. \label{eq:averagebe7error}
\end{equation}

\begin{figure}[!ht] \centerline{\psfig{file=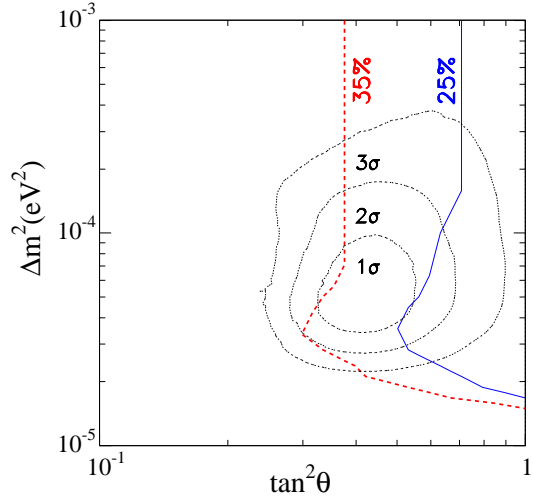,width=3in}}
\caption[]{Percentage error in determining the total $^7$Be flux.
The figure shows two contours (in \%) for the uncertainty in
determining the total $^7$Be flux. The uncertainties were
calculated as described in Sec.~\ref{subsec:be7procedure} and
Sec.~\ref{subsec:uncertaintiesbe7},  using a combined analysis
of SNO CC and gallium data together with simulated KamLAND data.
The curves labeled $1\sigma$, $2\sigma$, and $3\sigma$  represent
allowed regions from a global analysis of the available solar and
reactor data (see Fig.~\ref{fig:solarvskamlandreal}).}
\label{fig:be7errors}
\end{figure}

Figure~\ref{fig:be7errors} shows the contours of the maximum
percentage deviation (in absolute value) from the local best-fit
value of $f_{\rm Be}$. The figure shows that within the currently
allowed $1\sigma$ solar neutrino oscillation region the expected
uncertainty in the determination of the $^7$Be flux is of the
order of $25\%-35$\% [in agreement with
Eq.~(\ref{eq:averagebe7error})].

\subsection{Can one use the Chlorine experiment to determine the
$^7$Be solar neutrino flux?}\label{subsec:chlorine}

We can derive an expression for $f_{\rm Be}$ in terms of the
measured event rate in the
Chlorine~\cite{chlorine,chlorinehistorical} solar experiment.
Replacing ``gallium" by ``chlorine" everywhere in
Sec.~\ref{subsec:be7procedure}--Sec.~\ref{subsec:accuracybe7}, we
find\hfill\break
\vbox{
\begin{eqnarray}
f_{\rm Be}&=&\frac{1}{R^{\rm ^7Be,SSM}_{\rm Cl} \langle P_{ee}
(\Delta\bar m^2,\tan^2\bar\theta)_{\rm KamLAND} \rangle^{\rm
^7Be}_{\rm Cl}} \Big[R^{\rm exp}_{\rm Cl} - \sum_i R^{\rm
\phi_i,SSM}_{\rm Cl}
 \langle P_{ee} (\Delta\bar m^2,\tan^2\bar\theta)_{\rm KamLAND}
\rangle^{\phi_i}_{\rm Cl} \nonumber \\ [+0.2cm]  & & -R^{\rm
^8B,SSM}_{\rm Cl} \frac{R^{\rm CC,exp}_{\rm SNO}} {R^{\rm
CC,SSM}_{\rm SNO}}\frac{ \langle P_{ee} (\Delta\bar
m^2,\tan^2\bar\theta)_{\rm KamLAND} \rangle^{\rm ^8B}_{\rm Cl}}
{\langle P_{ee} (\Delta\bar m^2,\tan^2\bar\theta) _{\rm
KamLAND}\rangle_{\rm SNO}} \Big], \label{eq:fbechlorine}
\end{eqnarray}
}
where $R^{\rm exp}_{\rm Cl}=2.56\pm 0.23$ SNU~\cite{chlorine}.
Equation~(\ref{eq:fbechlorine}) is the analogue for the Chlorine
experiment of the previously derived Eq.~(\ref{eq:fbe}) that was
used in the discussion of extracting $f_{\rm Be}$ for the gallium
experiments.

Table~\ref{tab:fbecl} lists the best-fit values of $f_{\rm Be}$
that were obtained by solving Eq.~(\ref{eq:fbechlorine}) for
different pairs of oscillation parameters, $\left(\Delta
\bar{m}^2,\tan^2\bar{\theta}\right)$. The best-fit values of
$f_{\rm Be}$ obtained from the the Chlorine experiment are smaller
than the best-fit values inferred using the gallium data (cf.
Table~\ref{tab:fbe} and Table~\ref{tab:fbecl}). Using the chlorine
data, one can even get negative (i.e., unphysical) solutions for
$f_{\rm Be}$.

The basic reason for the difficulty in determining $f_{\rm Be}$ is
that the large expected rate from the $^8$B neutrino flux, even
after reduction due to neutrino oscillations, can account for all
the observed rate in the Chlorine experiment~\cite{bethe}. The
contribution from the $^7$Be neutrinos is obtained by subtracting
the large and somewhat uncertain expected contribution from $^8$B
neutrinos (and the contributions from CNO and $pep$ neutrinos that
are expected to be much less important) from the total measure
chlorine rate. The result is a rather small and uncertain
remainder, which can be attributed to $^7$Be neutrinos.

Table~\ref{tab:fbecl} also presents the most important sources of
uncertainty for inferring the value of $f_{\rm Be}$ using the
chlorine, rather than the gallium, data. Since the inferred
best-fit values for  $f_{\rm Be}$ that are obtained using the
chlorine data are very small (or even negative), we show in
Table~\ref{tab:fbecl}  the associated shifts in the prediction of
$f_{\rm Be}$. For the much more reliable inferences from gallium
data, we show instead in  Table~\ref{tab:fbe} the percentage
shifts, not the actual numerical shifts.
\begin{table}[!t]
\caption{Best-fit values of $f_{\rm Be}$ and associated
uncertainties (chlorine based simulation). This table is similar
to  Table~\ref{tab:fbe} except that for Table~\ref{tab:fbecl} data
from the Chlorine experiment were used instead of data from the
gallium experiment. Because the inferred values of $f_{\rm Be}$
are small and very uncertain using data from the Chlorine
experiment,  Table~\ref{tab:fbecl} presents uncertainties as the
numerical shift in the best-fit values (not as percentage
uncertainties, which are given in Table~\ref{tab:fbe}).
\label{tab:fbecl}}
\begin{tabular}{ccccccccc}
& & & \multicolumn{5}{c} {Uncertainties $\Delta(f_{\rm Be})_i$}\\
\noalign{\smallskip}\cline{4-9}\noalign{\smallskip}
$\Delta\bar m^2$&$\tan^2\bar\theta$& $f_{\rm Be}$ & ${\rm Cl, exp}$ & ${\rm Cl,
c.s.}$ & ${\rm SNO, exp}$ & ${\rm SNO, c.s.}$ & ${\rm CNO,flux}$  &
${\rm Total}$ \\
\noalign{\smallskip}\hline\noalign{\smallskip}
 $5.0\times10^{-5}$  &$4.2\times10^{-1}$
& 0.17& $0.37$  & $0.25$ &
$0.26$ & $0.07$  & $0.08$
& $0.50$ \\
\noalign{\medskip} $5.0\times10^{-5}$  &$5.01\times10^{-1}$ &
0.20& $0.38$ & $0.29$ &  $0.21$ & $0.07$  & $0.08$
& $0.53$ \\
\noalign{\medskip} $5.0\times10^{-5}$  &$2.51\times10^{-1}$ &
0.07& $0.32$ & $0.10$ &  $0.18$ & $0.06$  & $0.08$
& $0.39$ \\
\noalign{\medskip} $7.94\times10^{-5}$  &$4.2\times10^{-1}$ &
0.12& $0.36$  & $0.17$  &  $0.20$ & $0.07$  & $0.08$
& $0.46$ \\
\noalign{\medskip} $7.94\times10^{-5}$  &$5.01\times10^{-1}$ &
0.16& $0.38$  & $0.22$ & $0.21$ & $0.07$   & $0.08$
& $0.49$ \\
\noalign{\medskip} $7.94\times10^{-5}$  &$2.51\times10^{-1}$ &
$-0.10$& $0.31$  & $0.10$&  $0.17$ & $0.06$   & $0.08$
& $0.37$ \\
\noalign{\medskip} $3.16\times10^{-5}$  &$4.2\times10^{-1}$ &
0.25& $0.39$ & $0.34$  &
 $0.21$ & $0.07$  & $0.07$
& $0.57$ \\
\noalign{\medskip} $3.16\times10^{-5}$  &$5.01\times10^{-1}$ &
0.26& $0.40$ & $0.37$ &  $0.22$ & $0.07$   & $0.07$
& $0.60$ \\
\noalign{\medskip} $3.16\times10^{-5}$  &$2.51\times10^{-1}$ &
0.16& $0.36$ & $0.25$ &
 $0.18$ & $0.06$   & $0.07$
& $0.46$ \\
\end{tabular}
\end{table}

The largest uncertainties in determining $f_{\rm Be}$ using the
chlorine data are caused by the experimental errors in the
chlorine absorption rate ($\sim 0.35$) and the SNO CC absorption
rate ($\sim 0.25$). The uncertainty in the chlorine neutrino
absorption cross sections are also significant ($0.03$ to $0.35$,
depending upon the neutrino oscillation parameters). To be
explicit, the uncertainties in $f_{\rm Be}$ due to the theoretical
uncertainties in calculating the chlorine neutrino absorption
cross sections that were used in constructing
Table~\ref{tab:fbecl} were evaluated from the following equation,
\begin{equation}
\Delta(f_{\rm Be})^2_{\rm Cl,c.s.}= \left[\sum_{j=
pep,{\rm ^7Be,CNO}}(f_{\rm Be}({\rm c.s.}_j\pm 1 \sigma_{\rm
c.s._j})- f_{\rm Be}(i)\right]^2 + \left[\sum_{{\rm
j=  ^8B},hep}(f_{\rm Be}({\rm c.s.}_j\pm 1 \sigma_{\rm c.s._j}) -
f_{\rm Be}(i)\right]^2 \label{appex:chlorineferrors}~.
\end{equation}
(See the
Appendix for more details regarding the treatment of the cross
section errors.) Even the CNO fluxes ($\sim 0.08$) and the SNO CC
absorption cross section ($\sim 0.10$) contribute non-negligible
errors. The uncertainty from sterile neutrinos is small ($\sim
0.04$) and does not significantly affect the total uncertainty.

Remarkably, the individual uncertainties in $f_{\rm Be}$ which
arise from several different sources are larger than the current
best-fit values of $f_{\rm Be}$ (see Table~\ref{tab:fbecl}).

For the chlorine based determination, we have computed a
representative (average) shift in  $f_{\rm Be}$ by a Monte Carlo
sampling of the allowed KamLAND oscillation region shown in
Fig.~\ref{fig:solarvskamlandreal}. We find
\begin{equation}
\Delta f_{\rm Be} ~=~ 0.49, \, 1\sigma.
\label{eq:averagebe7shiftcl}
\end{equation}
This uncertainty is so large as to render not very useful the
determination of $f_{\rm Be}$ with the aid of chlorine data.

\section{How well can K\lowercase{am}LAND plus \boldmath$\nu-\lowercase{e}$ scattering experiments determine
the total $^8$B and $^7$B\lowercase{e} fluxes?} \label{sec:klandes}

 In this section, we show how data from KamLAND can be
combined with $\nu-e$ scattering data obtained with the
Super-Kamiokande and BOREXINO experiments in order to determine,
respectively, the total $^8$B (Sec.~\ref{subsec:kamlandsuperkb8})
and $^7$Be (Sec.~\ref{subsec:kamlandborexino}) solar neutrino
fluxes. As discussed before, the advantage of using purely CC
measurements to determine the fluxes is that the answers depend
only mildly on the active-sterile admixture. On the other hand,
$\nu-e$ scattering measurements have the advantage of smaller
uncertainties in the interaction cross sections.  Moreover, for
the determination of the $^7$Be flux, BOREXINO depends less
strongly on other neutrino fluxes predicted by the standard solar
model than do the radiochemical experiments, Chlorine, GALLEX,
SAGE, and GNO.

\subsection{How well can KamLAND and Super--Kamiokande determine
the total $^8$B flux?} \label{subsec:kamlandsuperkb8}

In this section, we show how data from the Super-Kamiokande and
KamLAND experiments can be combined to measure the total $^8$B
neutrino flux. We shall see that the determination using
Super-Kamiokande and KamLAND yields, on average, a  value for the
total flux than is comparable in precision to what is expected to
be obtained using SNO and KamLAND. The systematic uncertainties
are different in the two experiments, SNO and Super-Kamiokande.
Therefore, it will be important to compare the total $^8$B
neutrino flux that is inferred using Super-Kamiokande and KamLAND
with the value that is obtained using  SNO and KamLAND. If the
uncertainties are correctly estimated, then the two methods should
agree within their quoted errors.

The advantage of using purely charged current measurements with
SNO and KamLAND  to determine the total $^8$B flux is that the
answer depends only very slightly upon the unknown active-sterile
mixture (as discussed in Sec.~\ref{sec:determiningb8total}). The
principal advantages of the Super-Kamiokande experiment in the
present context is that the neutrino interaction cross section is
accurately known~\cite{nuecrosssection} and the statistical and
systematic errors have already been extensively
investigated~\cite{sksol00}. However, as we shall see below, when
using Super-Kamiokande there is a significant uncertainty ($+7$\%
and $-0$\%) in the total $^8$B flux due to the active-sterile
mixture, at least with our present knowledge of $\eta$.

The average survival probability for $^8$B solar neutrinos can be
written in the form

\begin{eqnarray}
\langle P(\Delta\bar m^2,\tan^2\bar\theta,\eta)_{\rm KamLAND}
\rangle_{\rm Super-Kamiokande}&=& \langle P_{ee}(\Delta\bar
m^2,\tan^2\bar\theta,\eta)_{\rm KamLAND} \rangle_{\rm
Super-Kamiokande}\nonumber\\&&+ \; \langle rP_{ex}(\Delta\bar
m^2,\tan^2\bar\theta,\eta)_{\rm KamLAND} \rangle_{\rm
Super-Kamiokande}, \;\label{eq:survivalscatteringsk}
\end{eqnarray}
where $P$ is the probability of oscillating into active neutrinos
and $r\equiv \sigma_{\mu}/\sigma_{e}\simeq 0.15$ is the ratio of
the the $\nu_e - e$ and $\nu_{\mu} - e$ elastic scattering
cross-sections. The expected sensitivity of $f_{\rm B}$ to the
principal sources of errors can be calculated from the following
equation:

\begin{equation}
\left(\frac{\sigma(f_{\rm B})} {f_{\rm B}}\right)^2 ~=~
\left(\frac{\sigma(R^{\rm exp}_{\rm Super-Kamiokande})} {R^{\rm
exp}_{\rm Super-Kamiokande}}\right)^2 + \left(\frac{\sigma(\langle
P\rangle_{\rm Super-Kamiokande,\,KamLAND})} {\langle P\rangle_{\rm
Super-Kamiokande,\,KamLAND})}\right)^2.
 \label{eq:errorfbosk}
\end{equation}
We suppose that KamLAND will observe (with associated
uncertainties that we simulate with a Monte Carlo) the rate
predicted for the global best-fit point shown in
Fig.~\ref{fig:solarvskamlandreal}. For purely active oscillations,
we find that
\begin{equation}
f_{\rm active~B}=1.07\left[1 ^{+0.037}_{-0.028} \,^{+0.040}_{-0.046} \right]=
1\left[1 \pm 0.054\right], ~~\cos^2 \eta = 1.0,
\label{eq:skfboactive}
\end{equation}
where the first error corresponds to the Super--Kamiokande
experimental uncertainty and the second error is caused by the
finite size of the allowed KamLAND region in oscillation parameter
space. The combined KamLAND and Super-Kamiokande measurement will,
for purely active neutrinos, yield a determination for the total
$^8$B flux that is more accurate than can be obtained with the SNO
CC measurement and KamLAND.
 Within the 1 $\sigma$ LMA region, the average uncertainty in
the  combined KamLAND and Super-Kamiokande measurement for the total active
$^8$B neutrino flux is expected to be
\begin{equation}
\sigma(f_{\rm active ~ B}) ~=~ 0.06 .
\label{eq:activesigmab8kamlandsk}
\end{equation}

For purely active neutrinos, Super-Kamiokande and KamLAND
may provide us with a more accurate determination of the total
$^8$B neutrino flux than SNO CC and KamLAND [cf.
Eq.~(\ref{eq:averagefbsigma}) and
Eq.~(\ref{eq:activesigmab8kamlandsk})].

Allowing for  the currently allowed $1\sigma$ ($25$\%) sterile
admixture, we find for the best fit point
\begin{equation}
f_{\rm total~B}=1.07\left[1 ^{+0.037}_{-0.028} \,^{+0.040}_{-0.046}
\;^{+0.07}_{-0.00}\right],~~\cos^2 \eta \geq 0.75
\label{eq:borexinofbosterile}
\end{equation}

The last error in Eq.~(\ref{eq:borexinofbosterile}) corresponds to
the present 1$\sigma$ uncertainty from the active-sterile
admixture $\cos^2\eta\geq 0.75$.
 The average uncertainty in
the  combined KamLAND and Super-Kamiokande measurement for
$f_{\rm total~B}$ is  expected to be
  $^{+9}_{-6}$\%,
within the 1 $\sigma$ LMA region and if the sterile admixture is as
large as currently allowed.

\subsection{How well can KamLAND and BOREXINO determine the total
$^7$Be flux?} \label{subsec:kamlandborexino}

In this section, we show how data from the KamLAND and BOREXINO
experiments can be combined to measure the total $^7$Be neutrino
flux.  The principal advantage of using the BOREXINO experiment
for this purpose is that the signal in the BOREXINO
experiment~\cite{borexino} is predicted to be dominated by $^7$Be
neutrinos, whereas $^7$Be solar neutrinos are expected to
contribute only a relatively small (and unlabeled) fraction to the
observed event rate in the gallium and chlorine radiochemical
measurements (see Table~\ref{tab:garates})\footnote{The BOREXINO
detector can measure the energy of the recoil electrons produced
by $\nu-e$ scattering. The radiochemical detectors do not have
energy resolution, only an energy threshold.}.

However, unlike the cases involving the gallium and chlorine
experiments that were discussed in Sec.~\ref{sec:be7}, which
include only $\nu_e$ (CC) absorption, the BOREXINO experiment
detects both $\nu_e-e$ scattering and $\nu_\mu-e$ and $\nu_\tau-e$
scattering. One must consider in the present case the extent to
which the active-sterile neutrino admixture influences the
detected event rate for each set of oscillation parameters,
$(\Delta\bar m^2,\tan\bar\theta^2)_{\rm KLAND}$, determined by
KamLAND. As we shall see quantitatively in the following
discussion, this uncertainty regarding the sterile admixture does
not decrease significantly the overall accuracy of the inferred
total $^7$Be neutrino flux.

The average survival probability for active neutrinos can be
written conveniently in the form

\begin{eqnarray}
\langle P(\Delta\bar m^2,\tan^2\bar\theta,\eta)_{\rm KamLAND}
\rangle^{\phi_i}_{\rm BOREXINO}&=& \langle P_{ee}(\Delta\bar
m^2,\tan^2\bar\theta,\eta)_{\rm KamLAND} \rangle^{\phi_i}_{\rm
BOREXINO}\nonumber\\&&+ \; \langle  rP_{ex}(\Delta\bar
m^2,\tan^2\bar\theta,\eta)_{\rm KamLAND} \rangle^{\phi_i}_{\rm
BOREXINO}, \;.\label{eq:survivalscattering}
\end{eqnarray}
where $P^{\phi_i}$ is the oscillation probability of neutrino
fluxes of source $\phi_i$ into active neutrinos and $r\equiv
\sigma_{\mu}/\sigma_{e}\simeq 0.15$ is the ratio of the the $\nu_e
- e$ and $\nu_{\mu} - e$ elastic scattering cross-sections.

The expression for $f_{\rm Be}$ has the same form for the case in
which the KamLAND and BOREXINO experiments are considered together
as for the previously discussed cases involving the gallium
experiments [see Eq.~(\ref{eq:fbe})] and the chlorine experiment
 [see Eq.~(\ref{eq:fbechlorine})]. Using the average survival
probability defined in Eq.~(\ref{eq:survivalscattering}), we can write

\begin{eqnarray}
f_{\rm Be}=\frac{1}{R^{\rm ^7Be,SSM}_{\rm BOREXINO}
\langle(P\Delta\bar m^2,\tan^2\bar\theta,\eta)_{\rm KamLAND}
\rangle^{\rm ^7Be}_{\rm BOREXINO}}
\Big[R^{\rm exp}_{\rm BOREXINO}\nonumber \\
- \sum_i R^{\rm \phi_i,SSM}_{\rm BOREXINO}
 \langle P(\Delta\bar m^2,\tan^2\bar\theta,\eta)_{\rm KamLAND}
\rangle^{\phi_i}_{\rm BOREXINO} \Big] \label{eq:fborexino}
\end{eqnarray}
In the last term of Eq.~(\ref{eq:fborexino}), we include the
contributions from $pp$, and CNO neutrinos. The contribution from
$^8$B neutrinos is negligible because the observed $^8$B neutrino
flux is about a factor of $10^3$ smaller than the predicted SSM
$^7$Be neutrino flux and because  $^8$B neutrinos primarily
produce high energy recoils electrons that the BOREXINO detector
can distinguish from the low energy recoil electrons produced by
$^7$Be neutrinos.

The expected sensitivity of $f_{\rm Be}$ to different sources of
errors is given by
\begin{equation}
\left(\frac{\sigma(f_{\rm Be})} {f_{\rm Be}}\right)^2 =
{\delta(f_{\rm Be})}^2_{\rm BOREXINO,exp} ~+~ {\delta(f_{\rm
Be})}^2_{\rm KamLAND}~+~ {\delta(f_{\rm Be})}^2_{\rm CNO} +~ {\rm
smaller~terms}. \label{eq:errorsborex}
\end{equation}

In order to estimate the accuracy of this method for determining
$f_{\rm Be}$, we suppose that KamLAND will observe (with associated
uncertainties that we simulate with a Monte Carlo) the rate
predicted for the global best-fit point\footnote{The predicted
rate in units of the expected SSM rate is
 $R^{\rm exp}_{\rm
BOREXINO}=0.64$, see Ref.~\cite{newbgp}.} shown in
Fig.~\ref{fig:solarvskamlandreal} and that BOREXINO also will
observe a signal corresponding to this best-fit point (with
associated uncertainties). In the absence of any published data,
we estimate that-based upon experience in previous solar neutrino
experiments~\cite{sno,sksol00,chlorine,sage,gallex,gno}-that
BOREXINO will achieve a systematic uncertainty of between $5$\%
and $10$\% . We estimate in this way that the combined experiments
will yield a determination of  $f_{\rm Be}$ that is, for purely
active oscillations,
\begin{equation}
f_{\rm active~Be}=1.00\left[1\pm 0.05 (0.1)\pm 0.020
\pm 0.020\right]= 1\left[1 \pm
{0.057} (\pm {0.103)}\right], ~~\cos^2 \eta =
1.0, \label{eq:borexinofbeactive}
\end{equation}
where the first error corresponds to the BOREXINO experimental
uncertainty, the second to the uncertainty in the reconstructed
KamLAND region, and the last error is due to the theoretical
uncertainty in the predicted CNO fluxes.
 Within the 1 $\sigma$ LMA region, the average uncertainty in
the  combined KamLAND and BOREXINO measurement for the total active
$^7$Be neutrino flux is expected to be
\begin{equation}
\sigma(f_{\rm active ~ Be}) ~=~ 0.06 (0.105)~.
\label{eq:activesigmab7kamlandbor}
\end{equation}
If we allow for a 25\%
sterile admixture, we find
\begin{equation}
f_{\rm total~Be}=1.00\left[1\pm 0.05 (0.1) \; \pm 0.02
\pm 0.020 \;^{+0.05}_{-0.00}\right] \label{eq:borexinofbesterile}~.
\end{equation}
The last error in Eq.~(\ref{eq:borexinofbesterile}) corresponds to
the present 1$\sigma$ uncertainty from the active-sterile
admixture  $\cos^2\eta\geq 0.75$. The numbers in parentheses in
Eq.~(\ref{eq:borexinofbeactive}) and
Eq.~(\ref{eq:borexinofbesterile}) correspond to assuming the
larger systematic uncertainty, $\pm 10$\%, for the BOREXINO
measured rate.

For $f_{\rm total~Be}$, the average uncertainty in
the  combined KamLAND and BOREXINO measurement is expected to be
$^{+8}_{-6}(^{+11}_{-10})$\%,
within the 1 $\sigma$ LMA region and  if the sterile admixture is as
large as currently allowed,

\section{Determination of  the \boldmath$\lowercase{pp}$ neutrino flux }
\label{sec:determiningpp}

In this section, we analyze three strategies for determining the
total $pp$ solar neutrino flux without requiring a dedicated
experiment that measures only the $pp$ neutrinos. We first
describe in Sec.~\ref{subsec:existingdata} how one can make a
crude determination of the $pp$ neutrino flux using the measured
Gallium, Chlorine, and SNO (CC) event
rates~\cite{sage2002,chlorine,gallex,gno} and the standard solar
model predictions for all but the $^8$B, $^7$Be, and $pp$ solar
neutrino fluxes. This part of the discussion is similar to the
analysis described in Ref.~\cite{sage2002}, although we evaluate
explicitly the uncertainty caused by the finite size of the
allowed region in oscillation parameter space . We then determine
in Sec.~\ref{subsec:galliumsno} how well one can infer the $pp$
flux using just the Gallium and the SNO measurements and the BP00
predictions for the other neutrino fluxes, especially the $^7$Be
neutrino flux. The unknown sterile-active mixture contributes only
a negligible error using the strategies described in
Sec.~\ref{subsec:existingdata} and Sec.~\ref{subsec:galliumsno}.
Next, we show in Sec.~\ref{subsec:ppkamlandborexino} how the
precision of the determination of the $pp$ flux can be improved by
using, in the future, the results from KamLAND (to constrain the
neutrino oscillation parameters) and from BOREXINO (to constrain
the $^7$Be neutrino flux).  The
principle of this strategy has
also been discussed by the SAGE collaboration~\cite{sage2002}.

The theoretical error on the $pp$ neutrino flux is $\pm 1$\% (see
Ref.~\cite{BP01}), which is an order of magnitude smaller than the
estimated errors that we find in this section on the empirically
derived values of $f_{pp}$.  To achieve a precision of $10$\% or
better in the determination of the solar neutrino $pp$ flux will
require a dedicated and accurate experiment devoted to measuring
the $pp$ flux.

Throughout this section, we treat the $pp$ flux and the $pep$
flux as a single variable because they are so closely linked
physically ~\cite{neutrinoastrophysics}.
The two reactions are linked because the $pep$
reaction is obtained from the $pp$ reaction by exchanging a positron in the
final state with an electron in the initial state. Thus the rates for the two
reactions are proportional to each other to high accuracy~\cite{pppep}.
For convenience, we
denote the sum of the $pp$ and $pep$ fluxes as simply $pp$.

\subsection{Using Chlorine, Gallium, and SNO data and BP00 predictions}
\label{subsec:existingdata}

In this subsection, we show how data from the chlorine, gallium,
and SNO experiments can be combined with the BP00 predictions for
the CNO and $hep$ fluxes to determine the total $pp$ neutrino flux.

The reduced $pp$ solar neutrino flux, $f_{pp}$ defined (by
analogy with $f_{\rm B}$) with respect to the predicted standard
solar model sum of the $pp$ and $pep$ fluxes, can be written as:
\begin{eqnarray}
f_{pp}&=&\frac{1}{R^{pp,{\rm SSM}}_{\rm Ga} \langle P_{ee}
(\Delta m^2,\tan^2\theta) \rangle^{pp}_{\rm Ga}}
\Big[R^{\rm exp}_{\rm Ga} - \sum_i f_i R^{\rm \phi_i,SSM}_{\rm Ga}
 \langle P_{ee} (\Delta m^2,\tan^2\theta)
\rangle^{\phi_i}_{\rm Ga}  \nonumber \\ [+0.2cm] & & - f_{\rm
Be} R^{\rm ^7Be,SSM}_{\rm Ga}
 \langle P_{ee}(\Delta m^2,\tan^2\theta)\rangle^{\rm ^7Be}_{\rm Ga}
-R^{\rm^8B,SSM}_{\rm Ga} \frac{R^{\rm CC,exp}_{\rm SNO}} {R^{\rm
CC,SSM}_{\rm SNO}} \frac{\langle P_{ee}(\Delta
m^2,\tan^2\theta)\rangle^{\rm ^8B}_{\rm Ga}} {\langle
P_{ee}(\Delta m^2,\tan^2\theta)\rangle^{\rm ^8B}_{\rm
SNO}} \Big] .\label{eq:fpp}
\end{eqnarray}
Here the sum over $i$ in Eq.~(\ref{eq:fpp}) refers to the three
CNO neutrino sources and the $hep$ neutrinos (see
Table~\ref{tab:garates}).

We insert in Eq.~(\ref{eq:fpp}) the expression for
 $f_{\rm Be}$
in terms of the Chlorine and SNO rates and the standard solar
model CNO and $hep$ neutrino fluxes. Explicitly,
\begin{eqnarray}
f_{\rm Be}&=&\frac{1}{R^{\rm ^7Be,SSM}_{\rm Cl} \langle P_{ee}
(\Delta m^2,\tan^2\theta)\rangle^{\rm
^7Be}_{\rm Cl}} \Big[R^{\rm exp}_{\rm Cl} - \sum_i f_i R^{\rm
\phi_i,SSM}_{\rm Cl}
 \langle P_{ee} (\Delta m^2,\tan^2\theta)
\rangle^{\phi_i}_{\rm Cl} \nonumber \\ [+0.2cm]  & & -R^{\rm
^8B,SSM}_{\rm Cl} \frac{R^{\rm CC,exp}_{\rm SNO}} {R^{\rm
CC,SSM}_{\rm SNO}}\frac{ \langle P_{ee} (\Delta
m^2,\tan^2\theta)\rangle^{\rm ^8B}_{\rm Cl}}
{\langle P_{ee} (\Delta m^2,\tan^2\theta)\rangle_{\rm SNO}}
\Big]. \label{eq:fppbechlorine}
\end{eqnarray}

For the special case of the best fit point in the LMA solution
region, we find
\begin{equation}
f_{pp}=1.41(1\pm 0.08 \,^{+0.03}_{-0.05} \,^{+0.009}_{-0.007}
\pm{0.06} \,^{+0.06}_{-0.19}\,^{+0.04}_{-0.11}) .
\label{eq:fppbestlma}
\end{equation}
The first error in Eq.~(\ref{eq:fppbestlma}), ($\pm 0.08$),
results from the weighted average  experimental error for the
Gallium experiment~\cite{sage2002,gallex,gno}. The second error,
($^{+0.03}_{-0.05}$), reflects the uncertainties in the calculated
neutrino absorption cross sections on gallium~\cite{GaCS}. The
third error, ($^{+0.009}_{-0.007}$), is caused by the
uncertainties in the calculated standard solar model CNO
fluxes~\cite{BP01}. The fourth error, ($\pm{0.06}$),  derives
from  the measurement errors in the  SNO CC experimental
data~\cite{snonc} and neutrino absorption cross
section~\cite{deuteriumcs}. The fifth error, ($^{+0.06}_{-0.19}$),
results from  the experimental error for the Chlorine event
rate~\cite{chlorine} and the sixth error, ($^{+0.04}_{-0.11}$),
reflects  the uncertainties in the calculated neutrino absorption
cross sections for
Chlorine~\cite{neutrinoastrophysics,bahcalllisietal}. The physical
constraint, $f_{\rm Be}\geq 0$, was imposed in obtaining these
errors.

Within the $1\sigma$ LMA allowed region, we find that the central
value of $f_{pp}$ can vary in the range  $1.32<f_{pp}<1.50$
and the best-fit value can be determined with an average
uncertainty of $^{+13}_{-24}$\%. The range of central values of
$f_{pp}$ always exceeds the BP00 value  because the observed
Chlorine rate implies a low value of $f_{\rm ^7Be}$. Since the
largest contributions to the Gallium rate are from $pp$ neutrino
and $^7$Be neutrinos, a low $^7$Be flux implies a relatively high
$pp$ flux.

We can summarize as follows the results of the simulations carried
out within the $1\sigma$ allowed LMA region for the $pp$ flux:
\begin{equation}
f_{pp}=1.41(1\pm 0.06 \,^{+0.13}_{-0.24}) = 1.41(1
^{+0.14}_{-0.25}). \label{eq:fppsummaryknown}
\end{equation}
The first error in Eq.~(\ref{eq:fppsummaryknown}) results from the
allowed range of neutrino oscillation parameters and the second
error results from  the uncertainties in all other recognized
sources of errors, combined quadratically. The result given in
Eq.~(\ref{eq:fppsummaryknown}) should  be compared with the quoted
estimate by the SAGE collaboration~\cite{sage2002}, $f_{pp}=
1.29(1\pm 0.23)$.  This agreement is very welcome since the SAGE
paper points out that they made ``Several approximations...whose
nature cannot be easily quantified."
\subsection{Using Gallium and SNO data and BP00 predictions}
\label{subsec:galliumsno}

In this subsection, we determine the range of allowed values for
$f_{pp}$ using the BP00 predictions (and uncertainties) for
the $^7$Be neutrinos as well as the CNO and $hep$ neutrinos. We use
in this subsection  data only from the Gallium and SNO solar
neutrino experiments.

Following the same line of reasoning as in
section~\ref{subsec:existingdata}, we find for the best fit point
in the LMA allowed oscillation region,
\begin{equation}
f_{pp}=1.05(1\pm 0.11 \,^{+0.05}_{-0.08} \,^{+0.02}_{-0.03}
\pm{0.007} \,^{+0.04}_{-0.02}). \label{eq:fbbestlmabe7}
\end{equation}
Just as in Eq.~(\ref{eq:fppbestlma}),  the first error in
Eq.~(\ref{eq:fbbestlmabe7}), ($\pm 0.11$), is the experimental
error from the weighted average event rate in the gallium
experiments, the second error,($^{+0.05}_{-0.08}$), is due to the
neutrino absorption cross section on gallium, the third error,
($^{+0.02}_{-0.03}$),  is due to the uncertainties in the BP00
predictions of the CNO neutrino fluxes, the fourth error,
($\pm{0.007}$), contains the uncertainty in the SNO CC
experimental rate and the calculated absorption cross sections on
deuterium.  The uncertainty in the BP00 prediction for the $^7$Be
neutrino flux contributes the  last error, ($^{+0.04}_{-0.02}$),
in Eq.~(\ref{eq:fbbestlmabe7}).

Within the
$1\sigma$ allowed LMA region,  we find the central value of
$f_{pp}$ varies in the range $0.93<f_{pp}<1.16$; the
best-fit value of  $f_{pp}$ can be determined with an average
uncertainty of $\pm 14$\%. We can summarize the determination of
$f_{pp}$ as follows:
\begin{equation}
 f_{pp}=1.05(1^{+0.105}_{-0.11} \pm
0.14)= 1.05(1\pm 0.18), \label{eq:fppsummarybe7}
\end{equation}
 where the first error is the uncertainty due to
the neutrino oscillation parameters and the second error contains
the uncertainties from all other sources.

\subsection{Using BOREXINO, KamLAND, Gallium and SNO data and BP00
predictions} \label{subsec:ppkamlandborexino}

In this subsection, we  show how the  uncertainty in $f_{pp}$
could be improved by using the KamLAND data to determine the
neutrino oscillation parameters and the BOREXINO  data, together
with the KamLAND oscillation parameters, to constrain $f_{\rm
^7Be}$. We start with Eq.~(\ref{eq:fpp}) but now use
 $f_{\rm Be}$ as determined from
BOREXINO and KamLAND data [Eq.~(\ref{eq:fborexino})].

We find that for the best-fit LMA solution (assuming that BOREXINO
finds the rate expected for this best fit point),
\begin{equation}
f_{pp}=1.05(1\pm 0.11 \,^{+0.05}_{-0.08} \,^{+0.01}_{-0.02}
\pm{0.007} \pm 0.05  \pm 0.02 (0.04) \,^{+0.00}_{-0.02})
\label{eq:fppkamlandborexino}
\end{equation}
The first error shown in Eq.~(\ref{eq:fppkamlandborexino}), ($\pm
0.11$), is the experimental error on the measured Gallium event
rate, the second error, ($^{+0.05}_{-0.08}$), represents the
uncertainties from the calculated gallium cross sections. The
third error, $(^{+0.01}_{-0.02})$,  is from the predicted CNO
fluxes and the fourth error, ($\pm{0.007}$), contains the
uncertainties from the SNO experimental data and calculated
deuterium absorption cross sections. The range of oscillation
parameters within the KamLAND reconstructed region gives the fifth
error, ($\pm 0.05$), and the sixth error, ($\pm 0.02$ [$\pm
0.04$]), results from  the uncertainty in the measured BOREXINO
event rate [which we take to be 5\% (10\%)]. The unknown
active-sterile admixture (which goes in the direction of lowering
$f_{pp}$) contributes the last error shown in
Eq.~(\ref{eq:fppkamlandborexino}).

On average, the the precision expected using the Gallium, KamLAND,
and SNO experiments  is
\begin{equation}
f_{pp} = 1.05[1.0 + 0.14(0.15)].
\label{eq:avfppborexinokamland}
\end{equation}
The dominant sources of error in
Eq.~(\ref{eq:avfppborexinokamland}) are the  experimental error in
the gallium rate and the uncertainties in the calculated gallium
absorption cross section.

\section{Discussion and Conclusions}
\label{sec:discuss}
 In this section, we review and discuss our principal conclusions. We
begin in Sec.~\ref{subsec:discusscurrentlyallowed} by summarizing
the experimentally-allowed range that currently exists for the
total $^8$B solar neutrino flux, taking account of the possibility
that sterile neutrinos exist. We then summarize in
Sec.~\ref{subsec:discussb8} how well the total $^8$B neutrino
flux, and separately the sterile $^8$B neutrino flux, can be
obtained by combining KamLAND and SNO measurements. Next we
describe in Sec.~\ref{subsec:discussbe7} how well the total $^7$Be
flux can be determined using data from the KamLAND, gallium
 (GALLEX, SAGE, and GNO), and SNO experiments, i.e., using only CC
disappearance experiments. In this same subsection, we summarize
the measurement accuracy that can be obtained using data from just
the KamLAND and the BOREXINO ($\nu-e$ scattering) experiments. We
describe three strategies (making use of existing and future solar
neutrino measurements and guidance from the BP00 solar model) for
determining the $pp$ solar neutrino flux without the help of an
experiment that measures separately the $pp$ neutrinos. We point
out in Sec.~\ref{subsec:discusscrosssections} (and in the
Appendix) that in order to determine the total $^8$B or $^7$Be
neutrino flux, and {\it a fortiori} to determine the sterile
contributions to these fluxes, the correlations among theoretical
errors for neutrino absorption cross sections and for solar
neutrino fluxes must be treated more accurately than in previous
discussions. Section~\ref{subsec:discussfocus} summarizes the main
focus of the present paper.

We concentrate  on procedures to determine experimentally the
total solar $^8$B neutrino flux and the total solar $^7$Be
neutrino flux in a universe in which sterile neutrinos might
exist. Our methods can work only if the LMA solution of the solar
neutrino problem is correct.

The numerical values given here for the precision with which
different quantities can be determined are obtained using
simulations of how well different experiments may perform.
Therefore, the numerical values are intended only as illustrative
guides to what may be possible.

\subsection{The currently allowed $^8$B solar neutrino flux if sterile
neutrinos exist} \label{subsec:discusscurrentlyallowed}

 The combined SNO CC data and the Super-Kamiokande
$\nu_e-e$ scattering data together yield a widely acclaimed
agreement between the $^8$B solar neutrino flux that is predicted
by the standard solar model and the empirically inferred $^8$B
neutrino flux. However, this agreement between solar model
prediction and solar neutrino measurement is not a unique
interpretation of the existing measurements if one allows for the
possibility that the incident solar neutrino flux could contain a
significant component of sterile neutrinos. We show in
Table~\ref{tab:fbranges} and in Sec.~\ref{sec:presentknowledge}
that if one takes account of the possibility that sterile
neutrinos may exist then the total solar $^8$B neutrino flux could
be as large as $2.3$ times the flux predicted by the standard
solar model.  In principle, the existing solar neutrino data could
be inconsistent with the standard solar model predictions.

\subsection{Determining the total $^8$B solar neutrino flux including sterile
neutrinos} \label{subsec:discussb8}

The total $^8$B solar neutrino flux, active plus sterile
neutrinos, can be determined with a typical $1\sigma$ accuracy of
about $10$\% by comparing the charged current measurements from
the KamLAND reactor experiment and the SNO experiments (see
Sec.~\ref{sec:determiningb8total}). The active $^8$B neutrino flux
has been measured by comparing the SNO CC flux (which measures
$\nu_e$) with the Super-Kamiokande neutrino-electron scattering
rate (which measures $\nu_e$ plus, with less sensitivity, $\nu_\mu
+ \nu_{\tau}$). The SNO neutral current measurement will provide
an additional and, ultimately, more accurate measurement of the
total active $^8$B solar neutrino flux.

By subtracting the independently-measured active $^8$B neutrino
flux from the total (active plus sterile) $^8$B neutrino flux, one
can determine empirically the sterile component of the solar
neutrino flux. The active $^8$B neutrino flux will eventually be
determined accurately by the SNO neutral current
measurement~\cite{sno,10com}. We estimate therefore that the
procedure described here has the potential of measuring, or
setting an upper limit on, the sterile component of the $^8$B
neutrino flux that is as small as $12$\% of the total $^8$B solar
neutrino flux.

The measurement of the total $^8$B neutrino flux, and the sterile
component of this flux, are independent of solar model
considerations. In order to establish the quantitative
conclusions, we have performed detailed simulations of the
accuracy of the KamLAND reactor experiment in determining neutrino
oscillation parameters (see Fig.~\ref{fig:solarvskamlandreal}) and
have evaluated the theoretical and experimental uncertainties that
affect the different flux determinations (see
Sec.~\ref{subsec:precisionb8} and the Appendix).

The combined measurements of the Super-Kamiokande and KamLAND
experiments can be used to determine independently a value for the
total $^8$B neutrino flux. This determination may be as accurate
as  $6$\% for purely
 active neutrinos. With the current limits on
the active-sterile admixture, the total $^8$B neutrino flux could
be inferred to an accuracy of $9$\% or better, as described in
Sec.~\ref{subsec:kamlandsuperkb8}. It will be important to compare
the value of the total $^8$B neutrino flux inferred by combining
the KamLAND and SNO charged current measurements with the value
obtained using the KamLAND and Super-Kamiokande experiments. This
comparison will be an important test of whether the systematic
uncertainties in the experiment and in the analyses are
understood.

\subsection{Determining the total $^7$Be solar neutrino flux
including sterile neutrinos} \label{subsec:discussbe7}

The total $^7$Be solar neutrino flux, active plus sterile
neutrinos, can be determined to a $1\sigma$  accuracy of about
$30$\% by combining measurements from KamLAND, SNO, and the
gallium experiments (see Sec.~\ref{sec:be7}). Unlike the purely
empirical determination that is possible for the $^8$B flux, the
measurement of the total $^7$Be solar neutrino flux requires some
assumption regarding the CNO solar neutrino fluxes. In our
estimates, we have assumed that the standard solar model
predictions for the CNO fluxes, and their uncertainties, are at
least approximately valid. Table~\ref{tab:fbe} shows that as long
as the CNO fluxes are not a factor of three or more larger than
the standard solar model predictions, then they will not
significantly limit the accuracy with which the total $^7$Be
neutrino flux can be determined. The measured capture rate in the
gallium experiments (GALLEX, SAGE, and GNO) currently constitutes
the largest recognized uncertainty in the determination of the
total $^7$Be flux by the method described here (see
Table~\ref{tab:fbe}). The constraints provided by the Chlorine
experiment are not very useful in providing an accurate
determination of the $^7$Be neutrino flux [see
Table~\ref{tab:fbecl} and Eq.~(\ref{eq:averagebe7shiftcl})].

One can also determine the allowed range of the total $^7$Be solar
neutrino flux using the data from the KamLAND reactor experiment
and the BOREXINO solar neutrino experiment. We show in
Sec.~\ref{subsec:kamlandborexino} that with this method one may
hope to obtain a $1\sigma$ accuracy of $11$\% or better for the
total $^7$Be solar neutrino flux, which is more accurate than we
estimate will be possible with the gallium and chlorine
radiochemical experiments.

 One can determine an upper
limit to the sterile component of the $^7$Be solar neutrino flux
by combining the measured rate in the neutrino-electron scattering
experiment, BOREXINO, with the $^7$Be total flux inferred from
measurements of the KamLAND, SNO, and gallium experiments. If one
assumes that the entire signal measured in BOREXINO is due to
$\nu_e$, then one obtains a minimum value for the active component
of the $^7$Be neutrino flux. Subtracting this minimum value from
the total $^7$Be flux, one will obtain an upper limit to the
sterile component of the flux. It seems unlikely that the
procedure described here has the sensitivity to measure a value
for the sterile component unless the sterile flux is larger than
$30$\% of the total $^7$Be neutrino flux. However, limits on the
sterile neutrino admixture can be obtained from  the analysis of
$^8$B and KamLAND neutrino measurements described in
Sec.~\ref{subsec:discussb8}.

In order to make a direct and precise measurement of the sterile
component of the $^7$Be solar neutrino flux, we need a charged
current measurement of the $^7$Be flux. A $^7$Be solar neutrino
absorption experiment, e.g., with a lithium target~\cite{lithium}
or with LENS~\cite{lens}, would make possible an accurate
determination of the sterile $^7$Be neutrino flux by providing a
set of experimental constraints that is analogous to what will
exist for the SNO, Super-Kamiokande, and KamLAND experiments.

\subsection{Determining the \boldmath$pp$ neutrino flux}
\label{subsec:discusspp}

The theoretical uncertainty in the calculated $pp$ solar neutrino
flux is estimated to be only $1\%$~\cite{BP01}.  Therefore, a
precise determination of the $pp$ solar neutrino flux will be of
great interest as a crucial test of the theory of stellar
evolution.  The measurement of the $pp$ neutrino flux will also
provide a critical test of whether the neutrino oscillation
theory, which works well at energies above $5$ MeV, also describes
accurately the lower energy neutrino phenomena (energies less than
$0.4$ MeV). We show in Sec.~\ref{sec:determiningpp} that the $pp$
flux can be determined with modest accuracy, of order $18$\% to
$20$\%, using a combination of existing experimental data and some
guidance from the BP00 standard solar model. In the future, it
should be possible to determine the $pp$ neutrino flux to an
accuracy of $15$\% using experimental data from the BOREXINO,
KamLAND, and SNO solar neutrino experiments and predictions (in a
non-critical way) from the standard solar model (see
Eq.~\ref{eq:avfppborexinokamland}). The unknown flux of sterile
neutrinos does not significantly affect the quoted estimates on
the accuracy with which the $pp$ flux can be determined.
To
measure the $pp$ flux with an accuracy sufficient to test
stringently the standard solar model prediction will require a
dedicated and accurate experiment that measures separately the
$pp$ neutrino flux.

\subsection{Correlation of errors, especially for neutrino
absorption cross sections} \label{subsec:discusscrosssections}

In order to determine the total $^8$B or $^7$Be solar neutrino
flux, and their sterile components, one must evaluate carefully
all known sources of error. In the course of this investigation,
we realized that the evaluations of the neutrino absorption cross
section uncertainties in previous neutrino oscillation studies,
including our own, have not properly taken account of the
correlations among the theoretical uncertainties in the cross
section calculations (for an insightful discussion of this point,
see Ref.~\cite{giuntics}). We discuss in the Appendix how the
cross section errors can be treated more correctly. We also
emphasize here that it is necessary to treat as fully correlated
the uncertainties in the principal CNO neutrino fluxes that are
obtained from standard solar model predictions. These effects are
small compared to other uncertainties in determining solar
neutrino oscillation parameters (the principal goal of nearly all
previous oscillation studies). The correlations among cross
section uncertainties become important only when one wants to make
accurate inferences regarding the neutrino fluxes themselves.

\subsection{Focus: total fluxes and sterile neutrino fluxes}
\label{subsec:discussfocus}
 The main focus in this paper is on determining experimentally the total
$^8$B, $^7$Be, and $pp$ solar neutrino fluxes in order to make
possible more precise tests of solar model predictions. We have
also shown that the contribution of sterile neutrinos to the total
flux can be measured, or a useful upper limit can be set, for
$^8$B solar neutrinos and for $^7$Be solar neutrinos.

\acknowledgments
  JNB acknowledges support from NSF grant No. PHY0070928.
 MCG-G is supported by the European Union Marie-Curie fellowship
 HPMF-CT-2000-00516.  This work was also supported by the Spanish
 DGICYT under grants PB98-0693  and  FPA2001-3031.

\appendix
\section*{}

We determine, as is conventional for many analyses of solar
neutrino data,  the allowed regions in the neutrino oscillation
space using a $\chi^2$ function that includes all the relevant
data. In the construction of the $\chi^2$ function, we have
followed closely the prescription of Ref.~\cite{fogli} (see also
Ref.~\cite{giuntics}), but we have included some modifications to
this prescription in order to account in more detail for the
energy dependence and the correlation of the cross section errors
for the Chlorine and gallium solar neutrino experiments.

For a given experiment $j$ (for example, gallium or chlorine), the
expected number of events can be written as
\begin{equation}
R_j~=~\sum_{i=1}^8 \sum_k \phi_i(E_k) C_{j}(E_k)
P_{ee,i}(E_k,\Delta m^2,\theta) ~\equiv~ \sum_{i=1}^8 R_{ij},
\label{eqapp:rate}
\end{equation}
where $i=1,8$ labels the solar neutrino fluxes, $\phi_i$, and $k$
labels the energy bins of energy $E_k$. The quantity $C_{j}(E_k)$
is the cross section for the interaction of a neutrino of energy
$E_k$ in the experiment $j$; $P_{ee,i}(E_k, \Delta m^2,\theta)$ is
the survival probability of $\nu_e$ for a given energy $E_k$ at a
specified  point, $\Delta m^2,\theta$, in neutrino oscillation
space\footnote{The survival probability depends upon the neutrino
species, $i$, because different neutrino species have different
probability distributions for the location of their production
within the Sun. For matter oscillations, the survival probability
obviously depends upon where in the Sun the neutrino was
produced.}.

The error matrix for the neutrino absorption cross sections can be
derived using Eq.~(\ref{eqapp:rate}). Since the errors are
uncorrelated  between different experiments, we need to evaluate
the following expression for a specified experiment:
\begin{equation}
\sigma^2({\rm c.s.})~=~\left<\left(R -R^0\right)^2\right>,
\label{appeq:defnsigma}
\end{equation}
where $R^0$ is best-estimate for the (gallium or chlorine)
experimental capture rate and the brackets indicate an average
over the probability distribution of uncertainties in the neutrino
cross sections.  Writing out the various terms in
Eq.~(\ref{appeq:defnsigma}), we find
\begin{equation}
\sigma^2({\rm c.s.})~=~\sum_{i,j} \sum_{k1,k2} \phi_i(E_{k1})
\phi_j(E_{k2})  P_{i}(E_{k1})P_j(E_{k2})<\Delta C(E_{k1})\Delta
C(E_{k2})>.
 \label{appeq:correlation}
\end{equation}
 In the standard Ref.~\cite{fogli}, the cross section error matrix for each experiment is given as:
\begin{eqnarray}
\sigma^2({\rm c.s.})_{jj}&~=~& \delta_{i1,i2} {\displaystyle
\sum_{i1} \sum_{i2}} R_{i1 j} R_{i2 j} \Delta \ln C_{i1}^j
\Delta \ln C_{i2}^j\nonumber  \\
&~=~& {\displaystyle}\sum_{i}  R^2_{i j}(\Delta\ln C_{i}^j)^2
\label{sigma2}
\end{eqnarray}
where $\Delta \ln C_{i}$ is the average uncertainty in the
interaction cross section of neutrinos of flux type $i$ in
experiment $j$.

Equation~(\ref{sigma2}) is obtained from Eq.~(\ref{eqapp:rate}) by
neglecting the correlations of the cross section errors among the
different neutrino fluxes and by neglecting the energy dependence
of the cross section errors. For the gallium detector, neither of
these assumptions is correct and even for chlorine the cross
section errors for different neutrino sources are strongly
correlated.

For all but the $^8$B and $hep$ neutrino fluxes, either ground
state to ground state transitions are the only energetically
possible transitions (which is the case for the chlorine detector)
or the ground state to ground state transitions dominate (which is
the case for the gallium detector). Thus all the cross sections
for the lower energy neutrinos move up or down together,
proportional to the square of the dominant matrix element.

When considering the gallium experiments, we include the energy
dependence of the cross section errors and assume full
correlations among the cross section errors of the different
neutrino sources that contribute at a specific energy.
Specifically, we use the results from Ref.~\cite{galliumcs} for
the cross section errors. For continuum sources, we take the
fractional error of the interaction cross section in the energy
bin $k$ to be
\begin{equation}
\Delta\ln C_{{\rm Ga,k}}~=~\frac{1}{3}\left[ \frac{|\sigma_{\rm
max,min}(E_k)-\sigma_{\rm best}(E_k)|} {\sigma_{\rm
best}(E_k)}\right],
\end{equation}
where $\sigma_{{\rm max,min}}(E_k)$ are the 3$\sigma$ upper and
lower limit cross sections given in Tables III and IV of
Ref.~\cite{galliumcs}. For the line sources, $^7$Be and $pep$, we
use the errors given in Eqs.~(41) and (42) of
Ref.~\cite{galliumcs}. To be complete, we have checked that the
shapes of these neutrino lines (see Ref.~\cite{be7shape}) do not
affect significantly the error estimates. The gallium contribution
to the cross section error matrix is therefore given by
\begin{equation}
\sigma_{{\rm Ga,Ga}}^2({\rm c.s.})~=~{\displaystyle \sum_{k1}
\sum_{k2}} \frac{\partial R}{\partial\ln C(E_{k1})} \frac{\partial
R}{\partial\ln C(E_{k2})} \Delta \ln C(k1) \Delta \ln C(k2) \;
\rho_{k1k2}, \label{sigma}
\end{equation}
where $\rho_{k1k2}$ is the correlation matrix for the cross
section errors of the different energy bins. For low energy
neutrino sources, the dominant process is the transition to the
germanium ground state and for higher energy neutrino sources
 ($E\gtorder 2$ MeV) transitions to excited states dominate.
Therefore, we assume the cross section errors to be fully
correlated between energy bins either below or above $E=2$ MeV
 ($\rho_{k1k2}=1$ for $E_{k1},E_{k2}< 2$ MeV or $E_{k1},E_{k2}> 2$
MeV). We take the errors to be uncorrelated between one energy bin
below $2$ MeV and one energy bin above $E=2$ MeV ($\rho_{k1k2}=0$
for $E_{k1}<2$ MeV and $E_{k2}> 2$ MeV, or $E_{k1}<2$ MeV and
$E_{k2}> 2$ MeV).

For chlorine, we take as fully correlated the cross section errors
for the $pp$, $pep$, $^7$Be, and CNO neutrinos. The cross
section errors for the $^8$B and $hep$ neutrinos are uncorrelated
with the errors for the lower energy neutrinos but are fully
correlated with each other. We can neglect for chlorine the energy
dependence of the chlorine to argon cross section errors because
the uncertainty for the lower energy neutrinos is determined
almost entirely from the ground-state to ground-state matrix
element. (Forbidden corrections are unimportant for these low
energy neutrinos.) For the $^8$B and $hep$ neutrinos, the
absorption cross sections are also dominated by a single (but
different) transition, in this case the super-allowed
transition~\cite{teo}, and therefore we can also neglect the
energy dependence for the higher energy neutrinos. We adopt the
values of the averaged chlorine cross section errors given in
Refs.~\cite{neutrinoastrophysics,bahcalllisietal}. To be explicit,
the chlorine contribution to the cross section error matrix is
given by
\begin{equation}
\begin{array}{ll}
\sigma_{\rm Cl,Cl}^2({\rm c.s.})= {\displaystyle \sum_{i1}
\sum_{i2}} R_{i1{\rm Cl}} R_{i2{\rm Cl}} \Delta \ln C^{\rm
Cl}_{i1}  \Delta \ln C^{\rm Cl}_{i2} \;\rho_{i1i2}
\label{eq:sigmaclcl}
\end{array}
\end{equation}
where $\rho_{i1i2}=1$ for $i1,i2=pep,{\rm CNO,^7Be}$ or
$i1,i2={\rm ^8B},hep$. Also, $\rho_{i1i2}=0$ for $i1=pep,{\rm
CNO,^7Be}$ and $i2={\rm ^8B},hep$.

\end{document}